\documentclass[12pt,english]{article}
\usepackage[T1]{fontenc}
\usepackage[latin1]{inputenc}
\usepackage{a4wide}
\usepackage{array}
\usepackage{amsmath}
\usepackage{graphicx}
\usepackage{amssymb}

\makeatletter

\providecommand{\tabularnewline}{\\}

 \newcommand{\lyxaddress}[1]{
   \par {\raggedright #1 
   \vspace{1.4em}
   \noindent\par}
 }

\usepackage{babel}
\makeatother
\begin{document}

\title{On the boundary form factor program}

\author{Z. Bajnok$^{1}$, L. Palla$^{2}$, and G. Takács$^{1}$}

\maketitle

\lyxaddress{\begin{center}$^{1}$\emph{\small Theoretical Physics Research Group,
Hungarian Academy of Sciences, }\\
\emph{\small 1117 Budapest, Pázmány Péter sétány 1/A, Hungary} \\
$^{2}$\emph{\small Institute for Theoretical Physics, Eötvös University,
}\\
\emph{\small 1117 Budapest, Pázmány Péter sétány 1/A, Hungary}\end{center}}

\begin{abstract}
Boundary form factor axioms are derived for the matrix elements of
local boundary operators in integrable 1+1 dimensional boundary quantum
field theories using the analyticity properties of correlators via
the boundary reduction formula. Minimal solutions are determined for
the integrable boundary perturbations of the free boson, free fermion
(Ising), Lee-Yang and sinh-Gordon models and the two point functions
calculated from them are checked against the exact solutions in the
free cases and against the conformal data in the ultraviolet limit
for the Lee-Yang model. In the case of the free boson/fermion the
dimension of the solution space of the boundary form factor equation
is shown to match the number of independent local operators. We obtain
excellent agreement which proves not only the correctness of the solutions
but also confirms the form factor axioms. 
\end{abstract}

\section{Introduction}

The bootstrap program aims to classify and explicitly solve 1+1 dimensional
integrable quantum field theories by constructing all of their Wightman
functions. The first stage is the S-matrix bootstrap: the scattering
matrix, connecting asymptotic \emph{in} and \emph{out} states, is
determined from its properties such as factorizability, unitarity,
crossing symmetry and the Yang-Baxter equation (YBE) supplemented
by the maximal analyticity assumption. The result is the complete
on-shell solution of the theory, i.e. the spectrum of excitations
and their scattering amplitudes, which can be related to some independent
definition of the model as a perturbed conformal field theory or a
Lagrangian QFT (for reviews see \cite{Sbstr,Sanal}). The second step
is the form factor bootstrap, which allows one to determine matrix
elements of local operators between asymptotic states using their
analytic properties originating from the already known S-matrix. Supposing
maximal analyticity leads to a set of solutions each of which corresponds
to a local operator of the theory. The form factors are then used
to build the correlation (Wightman) functions via their spectral representations,
yielding a complete off-shell description of the theory (see \cite{Smirnov,Arndt}
for reviews). 

The first step of an analogous bootstrap program for 1+1 dimensional
integrable \emph{boundary} quantum field theories, the boundary R-matrix
bootstrap, has been developed for several theories. In boundary theories
the asymptotic states are connected by the reflection R-matrix, which
satisfies unitarity and boundary crossing unitarity; for integrable
boundary QFT, it also satisfies the boundary YBE (BYBE) and boundary
bootstrap requirements. These equations supplemented by maximal analyticity
assumptions make possible to determine the reflection matrices and
provide the complete information about the theory on the mass shell
\cite{GZ}. 

For the second step matrix elements of local operators between asymptotic
states have to be computed. In a boundary quantum field theory there
are two types of operators, the bulk and the boundary operators, where
their names indicate their localization point. Due to the broken translational
invariance one point functions of bulk operators may acquire nontrivial
space dependence behaving analogously to the two point functions in
a bulk theory. Indeed this one point function can be calculated in
the crossed channel, where the role of time and space is changed and
the spatial boundary appears as a temporal one represented as an initial
(boundary) state in the matrix element. Inserting a complete system
of the bulk Hilbert space a spectral representation for the one point
functions can be obtained in terms of the bulk form factors and the
matrix element of the boundary state \cite{Muss1pt,DPTW1pt}. Truncating
this expansion at finite intermediate states provides a convergent
large distance expansion. However, matrix elements of boundary operators
cannot be computed in this way and the purpose of the present paper
is to develop a technique to compute their correlation functions. 

In this paper we initiate the second step of the boundary bootstrap
program, namely the boundary form factor program for calculating the
matrix elements of local boundary operators between asymptotic states.
We derive their analytic structure from that of the R-matrix which,
when supplemented by the assumption of maximal analyticity, leads
to their determination. In the bulk case, it was shown in \cite{cardy_mussardo}
that the solution space of the form factor equations can be brought
into one-to-one correspondence with the operator content of the model.
Based on this, we expect that the classification of the solutions
of the boundary form factor axioms provides information on the boundary
operator content of the theory, which in the ultraviolet limit is
in a one-to-one correspondence with the Hilbert space of the model.
Using the explicit form of the boundary form factors the spectral
representation for the boundary correlation functions can be obtained.

The paper is organized as follows: first we define the boundary form
factors by introducing asymptotic \emph{in} and \emph{out} multi-particle
states, which are related by the multi-particle reflection matrix.
Simple crossing relations are presented from which the form factor
axioms follow easily, and then the axioms are verified by some consistency
requirements. We outline a general strategy to solve theories with
diagonal bulk scattering and boundary reflection amplitudes, and to
compare the resulting two-point functions with their ultraviolet limits.
This idea is applied to integrable boundary perturbations of several
models, such as the free boson model, free fermion (alias Ising) field
theory, the scaling Lee-Yang model and sinh-Gordon theory. Appendix
A contains a heuristic derivation of the crossing relations from the
boundary reduction formula \cite{BBT}, while in Appendix B we present
a formal derivation of the boundary form factor axioms from the boundary
version of the Faddeev-Zamolodchikov algebra.

\section{Boundary form factors}

\subsection{Definitions}

The Hilbert space of a boundary quantum field theory consists of multi-particle
states, which can be labeled by the particle species and the corresponding
particle energies. To simplify the notations we restrict ourselves
to theories containing only one particle type with a given mass $m$.
In 1+1 dimensions it is convenient to work with the rapidity variable
$\theta_{i}$; the energy $E_{i}$ of the particle can be written
as $E_{i}=m\cosh\theta_{i}$, while the momentum is $p_{i}=m\sinh\theta_{i}$.
Following the evolution of the multi-particle state in time to $t\to-\infty$
the particles get far away form each other and from the boundary,
therefore forming an \emph{in} state which is equivalent to a free
multi-particle state and is denoted as%
\footnote{In general, particles in an interacting two dimensional quantum field
theory have an effective fermionic statistics with the sole exception
of free bosonic theories, for which it is necessary to allow equality
in the ordering of the particle rapidities.%
} \[
\vert\theta_{1},\theta_{2},\dots,\theta_{n}\rangle_{in}\quad;\qquad\theta_{1}>\theta_{2}>\dots>\theta_{n}>0\]
Positivity of all incoming rapidities is a consequence of the assumption
that the boundary is at the right end of the half line and it is a
major difference from the bulk situation. This difference is essential
because it influences the analyticity domain of matrix elements. 

For $t\to+\infty$ all the scatterings and reflections are terminated,
the particles are again far away from each other and from the boundary
forming the \emph{out} state, 

\[
\vert\theta_{1}^{'},\theta_{2}^{'},\dots,\theta_{m}^{'}\rangle_{out}\quad;\qquad\theta_{1}^{'}<\theta_{2}^{'}<\dots<\theta_{m}^{'}<0\]
which is again equivalent to a free state. By the standard assumption
of asymptotic completeness, the two sets of states form a complete
basis separately and are connected by the reflection matrix, which
is the boundary analogue of the S matrix. In an integrable theory,
due to the infinite number of conserved charges, there is no particle
creation ($n=m$), the set of rapidities changes only sign $\theta_{i}^{'}=-\theta_{i}$,
and the reflection matrix factorizes into the product of pairwise
bulk scatterings and individual reflections \begin{equation}
\vert\theta_{1},\theta_{2},\dots,\theta_{n}\rangle_{in}=\prod_{i<j}S(\theta_{i}-\theta_{j})S(\theta_{i}+\theta_{j})\prod_{i}R(\theta_{i})\vert-\theta_{1},-\theta_{2},\dots,-\theta_{n}\rangle_{out}\label{Rmatrix}\end{equation}
where $S(\theta_{i}-\theta_{j})$ connects the two particle asymptotic
\emph{in} and \emph{out} states in the bulk theory (without the boundary)

\begin{tabular}{cc}
\begin{tabular}{c}
$\vert\theta_{1},\theta_{2}\rangle_{in}^{bulk}=S(\theta_{1}-\theta_{2})\vert\theta_{2},\theta_{1}\rangle_{out}^{bulk}\qquad\qquad$depicted
as\tabularnewline
\tabularnewline
\end{tabular}&
\begin{tabular}{c}
\tabularnewline
\includegraphics[%
  height=2cm]{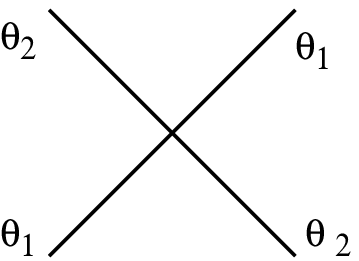}\tabularnewline
\end{tabular}\tabularnewline
\end{tabular}

It is defined originally for $\theta_{1}>\theta_{2}$ but can be analytically
continued for complex rapidity parameters such that the extended function
(denoted the same way) is meromorphic and satisfies unitarity and
crossing symmetry \[
S(\theta)S(-\theta)=1\quad,\qquad S(i\pi-\theta)=S(\theta)\]
It might have poles on the imaginary axis at locations $\theta=iu_{j}$
with residue $-i\textrm{res}_{\theta=iu_{j}}S(\theta)=\Gamma_{j}^{2}$,
some of which correspond to bound states. 

The amplitude $R(\theta)$ connects the one particle asymptotic states
in the boundary theory 

\begin{tabular}{cc}
\begin{tabular}{c}
$\vert\theta\rangle_{in}=R(\theta)\vert-\theta\rangle_{out}\qquad\qquad$depicted
as \tabularnewline
\tabularnewline
\end{tabular}&
\begin{tabular}{c}
\tabularnewline
\hspace{2cm}\includegraphics[%
  height=2cm]{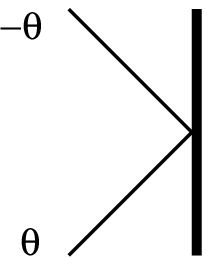}\tabularnewline
\end{tabular}\tabularnewline
\end{tabular}

It can also be extended from the fundamental domain $\theta>0$ to
a meromorphic function on the whole complex $\theta$ plane satisfying
unitarity and boundary crossing unitarity \[
R(\theta)R(-\theta)=1\quad,\qquad R(i\pi-\theta)S(2\theta)=R(\theta)\]
$R(\theta)$ may have poles at imaginary locations $\theta=iv_{j}$
($0<v_{j}<\pi/2$), some corresponding to excited boundary states.
If the interpolating field has a nontrivial vacuum expectation value
then generally there is also a pole at $\theta=i\pi/2$ with residue
\begin{equation}
-i\mathop{\textrm{Res}}_{\theta=\frac{i\pi}{2}}R(\theta)=\frac{g^{2}}{2}\label{eq:Defg}\end{equation}

The boundary form factor is defined as the matrix element of some
local boundary operator, $\mathcal{O}(t)$, between asymptotic states
\begin{eqnarray*}
\,_{out}\langle\theta_{1}^{'},\theta_{2}^{'},\dots,\theta_{m}^{'}\vert\mathcal{O}(t)\vert\theta_{1},\theta_{2},\dots,\theta_{n}\rangle_{in} & =\\
 &  & \hspace{-2cm}F_{mn}^{\mathcal{O}}(\theta_{1}^{'},\theta_{2}^{'},\dots,\theta_{m}^{'};\theta_{1},\theta_{2},\dots,\theta_{n})e^{-imt(\sum\cosh\theta_{i}-\sum\cosh\theta_{j}^{'})}\end{eqnarray*}
These form factors are defined only for $\theta_{1}>\theta_{2}>\dots>\theta_{n}>0$
and $\theta_{1}^{'}<\theta_{2}^{'}<\dots<\theta_{m}^{'}<0$. We can
introduce other form factors as \[
\,_{out}\langle\theta_{1}^{'},\theta_{2}^{'},\dots,\theta_{m}^{'}\vert\mathcal{O}(t)\vert-\theta_{1},-\theta_{2},\dots,-\theta_{n}\rangle_{out}=F_{mn}^{\mathcal{O}}(\theta_{1}^{'},\theta_{2}^{'},\dots,\theta_{m}^{'};-\theta_{1},-\theta_{2},\dots,-\theta_{n})\]
 and consider them as a continuation of the original ones in the rapidities.
Expressing these form factors (via the boundary reduction formula
\cite{BBT}) in terms of correlation functions an analytic continuation
can be performed for any (even) complex values of the rapidity parameters.
As a result the generalized form factors are meromorphic functions
of the rapidity parameters, and we shall assume that their poles always
have physical origins (maximal analyticity assumption). From the crossing
formula \begin{equation}
F_{mn}^{\mathcal{O}}(\theta_{1}^{'},\theta_{2}^{'},\dots,\theta_{m}^{'};\theta_{1},\theta_{2},\dots,\theta_{n})=F_{m-1n+1}^{\mathcal{O}}(\theta_{2}^{'},\dots,\theta_{m}^{'};\theta_{1}^{'}+i\pi,\theta_{1},\theta_{2},\dots,\theta_{n})+\textrm{disc}\label{eq:crossing}\end{equation}
derived in Appendix A, we can express all the form factors in terms
of the elementary form factors\[
\,_{out}\langle0\vert\mathcal{O}(0)\vert\theta_{1},\theta_{2},\dots,\theta_{n}\rangle_{in}=F_{n}^{\mathcal{O}}(\theta_{1},\theta_{2},\dots,\theta_{n})\]
It is important to notice that the boundary form factors $F_{n}^{\mathcal{O}}(\theta_{1},\dots,\theta_{n})$,
in contrast to the bulk case, do depend in general on all the rapidities
$\theta_{i}$, not just on their differences, since in the presence
of a boundary Lorentz invariance is broken.

\subsection{Axioms}

In the Appendices we derive all the following properties of the matrix
elements of local boundary operators valid in any integrable boundary
quantum field theory. Following the general philosophy in the bulk
case \cite{Smirnov} we take them as axioms defining the local operators
via their matrix elements. 

I. Permutation:

\begin{center}\[
F_{n}^{\mathcal{O}}(\theta_{1},\dots,\theta_{i},\theta_{i+1},\dots,\theta_{n})=S(\theta_{i}-\theta_{i+1})F_{n}^{\mathcal{O}}(\theta_{1},\dots,\theta_{i+1},\theta_{i},\dots,\theta_{n})\]
\includegraphics[%
  height=3cm,
  keepaspectratio]{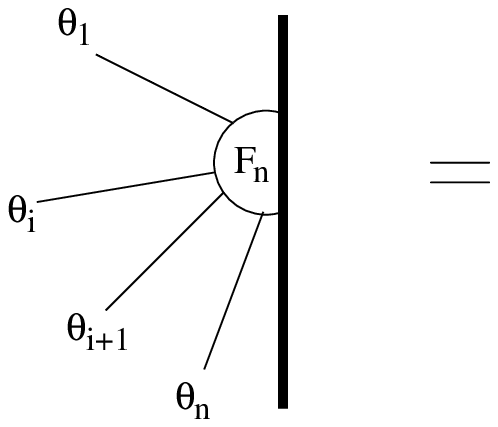}~~~\includegraphics[%
  height=3cm,
  keepaspectratio]{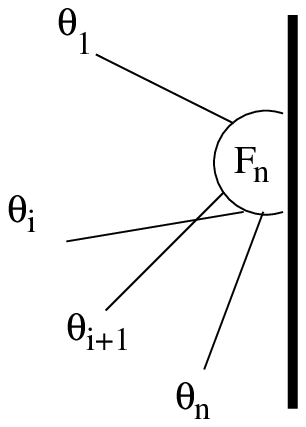}\end{center}

II. Reflection:\[
F_{n}^{\mathcal{O}}(\theta_{1},\dots,\theta_{n-1},\theta_{n})=R(\theta_{n})F_{n}^{\mathcal{O}}(\theta_{1},\dots,\theta_{n-1},-\theta_{n})\]

\begin{center}\includegraphics[%
  height=3cm,
  keepaspectratio]{bd1.eps}~~~ \includegraphics[%
  height=33mm,
  keepaspectratio]{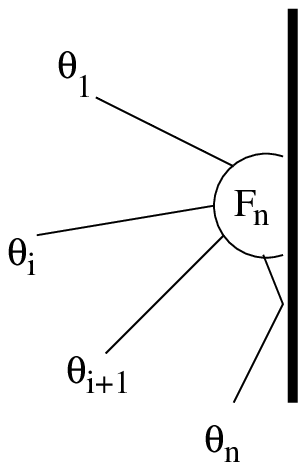}\end{center}

III. Crossing reflection: \[
F_{n}^{\mathcal{O}}(\theta_{1},\theta_{2},\dots,\theta_{n})=R(i\pi-\theta_{1})F_{n}^{\mathcal{O}}(2i\pi-\theta_{1},\theta_{2},\dots,\theta_{n})\]

\begin{center}\includegraphics[%
  height=3cm,
  keepaspectratio]{bd1.eps}~~~\includegraphics[%
  height=3cm,
  keepaspectratio]{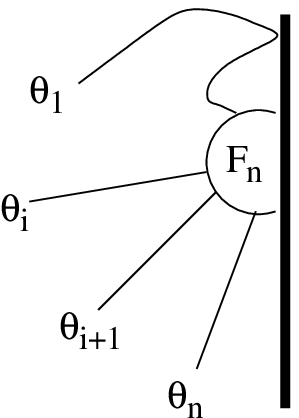}\end{center}

The singularity structure of the form factors is determined on physical
grounds and can be axiomatized as follows:

IV. Kinematical singularity\[
-i\mathop{\textrm{Res}}_{\theta=\theta^{'}}F_{n+2}^{\mathcal{O}}(\theta+i\pi,\theta^{'},\theta_{1},\dots,\theta_{n})=\left(1-\prod_{i=1}^{n}S(\theta-\theta_{i})S(\theta+\theta_{i})\right)F_{n}^{\mathcal{O}}(\theta_{1},\dots,\theta_{n})\]
or equivalently described as \[
-i\mathop{\textrm{Res}}_{\theta=\theta^{'}}F_{n+2}^{\mathcal{O}}(-\theta+i\pi,\theta^{'},\theta_{1},\dots,\theta_{n})=\left(R(\theta)-\prod_{i=1}^{n}S(\theta-\theta_{i})R(\theta)S(\theta+\theta_{i})\right)F_{n}^{\mathcal{O}}(\theta_{1},\dots,\theta_{n})\]

\begin{center}\includegraphics[%
  height=3cm,
  keepaspectratio]{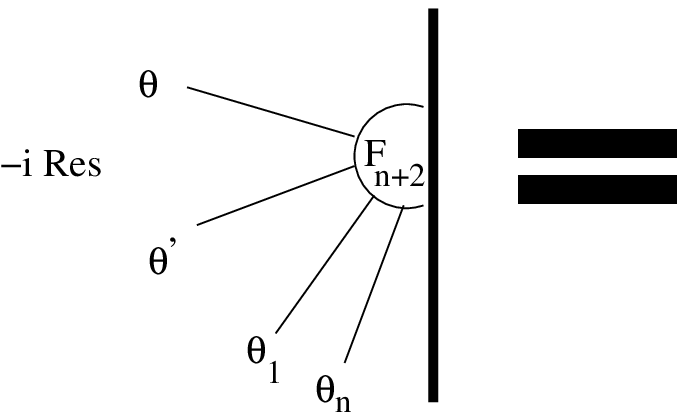}~~~\includegraphics[%
  height=3cm,
  keepaspectratio]{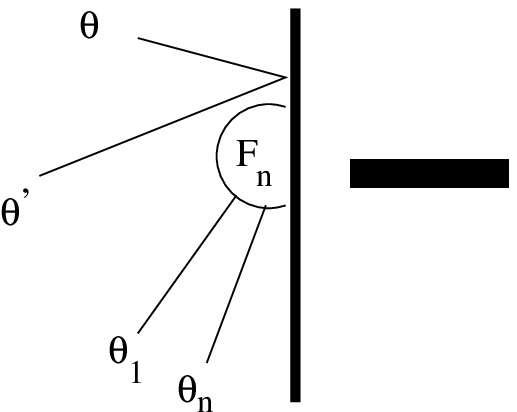}\includegraphics[%
  height=3cm,
  keepaspectratio]{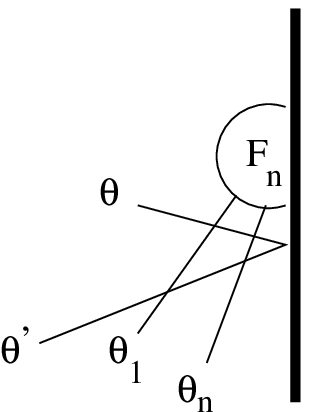}\end{center}

V. Boundary kinematical singularity

\[
-i\mathop{\textrm{Res}}_{\theta=0}F_{n+1}^{\mathcal{O}}(\theta+\frac{i\pi}{2},\theta_{1},\dots,\theta_{n})=\frac{g}{2}\Bigl(1-\prod_{i=1}^{n}S\bigl(\frac{i\pi}{2}-\theta_{i}\bigr)\Bigr)F_{n}^{\mathcal{O}}(\theta_{1},\dots,\theta_{n})\]

\begin{center}\includegraphics[%
  height=3cm,
  keepaspectratio]{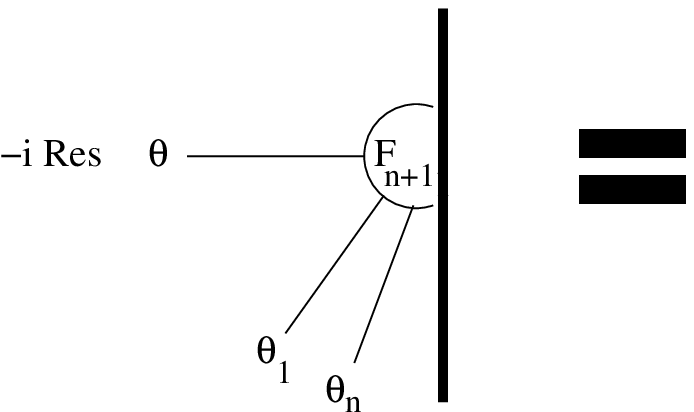}~~~\includegraphics[%
  height=3cm,
  keepaspectratio]{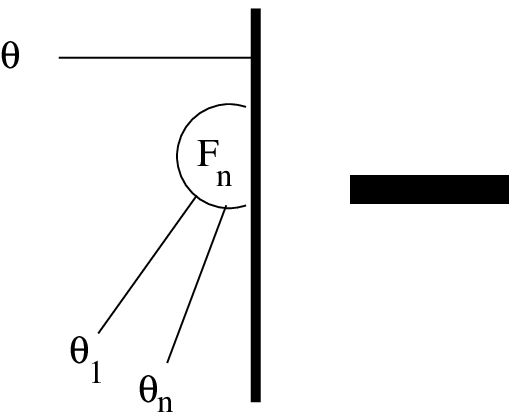}\includegraphics[%
  height=3cm,
  keepaspectratio]{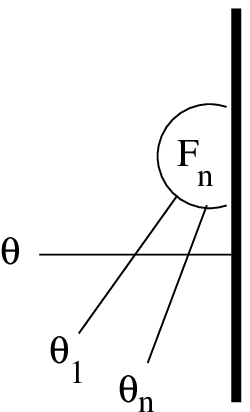}\end{center}

VI. Bulk dynamical singularity \[
-i\mathop{\textrm{Res}}_{\theta=\theta^{'}}F_{n+2}^{\mathcal{O}}(\theta+iu,\theta^{'}-iu,\theta_{1},\dots,\theta_{n})=\Gamma F_{n+1}^{\mathcal{O}}(\theta,\theta_{1},\dots,\theta_{n})\]

\begin{center}\includegraphics[%
  height=3cm,
  keepaspectratio]{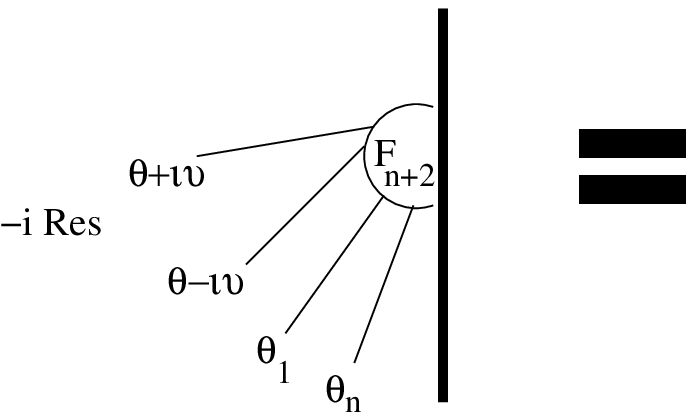}~~~\includegraphics[%
  height=3cm,
  keepaspectratio]{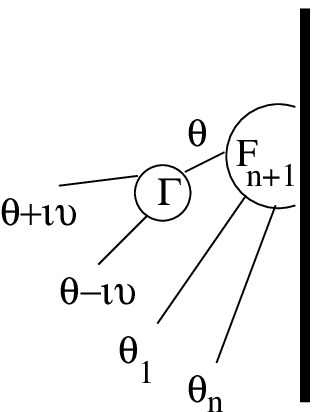}\end{center}

VII. Boundary dynamical singularity

\[
-i\mathop{\textrm{Res}}_{\theta=iv}F_{n+1}^{\mathcal{O}}(\theta_{1},\dots,\theta_{n},\theta)=\tilde{g}\tilde{F}^{\mathcal{O}}(\theta_{1},\dots,\theta_{n})\]

.

\begin{center}\includegraphics[%
  height=3cm,
  keepaspectratio]{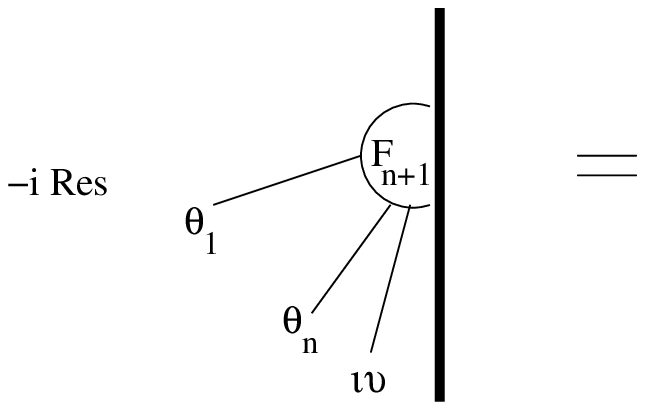}~~~\includegraphics[%
  height=3cm,
  keepaspectratio]{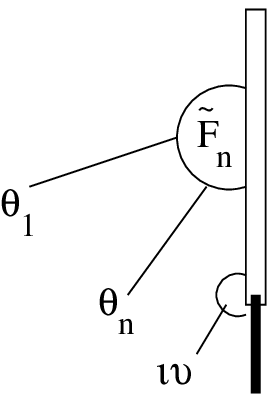}\end{center}

We note that equations similar to some of ours have been obtained
earlier studying boundary form factors in specific spin chains. Using
a concrete realization for the Hilbert space and the operators, these
equations were extracted originally for the XXZ and XYZ models in
\cite{XXZ} and extended for other spin chains in \cite{XXZGen}.
By extending the bulk free field representation for the boundary sine-Gordon
model the analogues of XXZ equations were obtained in \cite{SGff}.
In all these approaches, however, there is no analogue of the axiom
V, without which the equations do not determine completely the form
factors as can be seen on the example of the sinh-Gordon model. In
contrast, in our approach the form factor axioms are firmly established
from first principles of local quantum field theory, thus they are
valid in a general setting. As a further result of our systematic
approach the axioms found form a complete system ready to be solved.

\subsection{Consistency checks}

Before proceeding to concrete examples we perform a few consistency
checks of the axioms. First we note that they are self-consistent
in the sense that for specific rapidities the $n+2$ particle form
factor can be connected to the $n$ particle form factor either by
the kinematical singularity equations or by using twice the boundary
kinematical equations, and the two procedures give the same result.
Indeed taking double residue in the first case, first at $\theta=\theta^{'}$
and then at $\theta=i\frac{\pi}{2}$ gives \begin{eqnarray*}
i\mathop{\textrm{Res}}_{\theta=\frac{i\pi}{2}}i\mathop{\textrm{Res}}_{\theta^{'}=\theta}F_{n+2}^{\mathcal{O}}(-\theta+i\pi,\theta^{'},\theta_{1},\dots,\theta_{n}) & =\\
 &  & \hspace{-4cm}\left(-i\mathop{\textrm{Res}}_{\theta=\frac{i\pi}{2}}R(\theta)\right)\left(1-\prod_{i=1}^{n}S(\frac{i\pi}{2}-\theta_{i})S(\frac{i\pi}{2}+\theta_{i})\right)F_{n}^{\mathcal{O}}(\theta_{1},\dots,\theta_{n})\end{eqnarray*}
Taking now the residue at $\theta=\frac{i\pi}{2}$ first then at $\theta^{'}=\frac{i\pi}{2}$
and using that $S(0)=-1$ gives \begin{eqnarray*}
i\mathop{\textrm{Res}}_{\theta=\frac{i\pi}{2}}i\mathop{\textrm{Res}}_{\theta^{'}=\frac{i\pi}{2}}F_{n+2}^{\mathcal{O}}(-\theta+i\pi,\theta^{'},\theta_{1},\dots,\theta_{n}) & =\\
 &  & \hspace{-4cm}\frac{g^{2}}{4}\left(1+\prod_{i=1}^{n}S(\frac{i\pi}{2}-\theta_{i})\right)\left(1-\prod_{i=1}^{n}S(\frac{i\pi}{2}-\theta_{i})\right)F_{n}^{\mathcal{O}}(\theta_{1},\dots,\theta_{n})\end{eqnarray*}
 The two different orders of taking the residues differ by a factor
of $2$ since in the first case after taking the residue at $\theta^{'}=\theta$
we get a factor $f(2\theta-i\pi)$ which has a zero at $\theta=i\frac{\pi}{2}$
(due to $S(0)=-1$ the bulk minimal form factor vanishes at the threshold:
$f(0)=0$). In the second case after taking the first residue a factor
$f(\theta-i\frac{\pi}{2})$ appears. When expanding around $\theta=i\frac{\pi}{2}$
to take the second residue there appears a factor $2$ due to the
difference in the arguments of this particular factor (all other terms
are identical in the two cases). Combining the crossing symmetry of
the S-matrix with the definition of $g$ (\ref{eq:Defg}) the two
expressions are easily seen to be equivalent.

It is worth emphasizing that in the boundary kinematical singularity
axiom it is the particle-boundary coupling constant $g$ which appears
although the residue of the reflection factor determines only $g^{2}$.
There are known examples where in two physically different situations
the fundamental reflection amplitudes are the same and the two cases
are distinguished only by the sign of $g$ (e.g. the boundary Lee-Yang
model with $1$ boundary and with $\Phi$ boundary with a particular
value of the boundary coupling \cite{DTWbct} -- see in more details
in Sec. 3.3). Because of axiom V the solutions of the form factor
axioms are different for the two cases, as shown in detail in Sec.
3.3.

As a second test we relate the two disconnected physical domains (\emph{in/out})
of the definition of the form factor. By permuting successively each
rapidity to the last position, applying a reflection and permuting
back to their original position we obtain that \[
F_{n}^{\mathcal{O}}(\theta_{1},\dots,\theta_{n})=\prod_{i<j}S(\theta_{i}+\theta_{j})\prod_{i}R(\theta_{i})\prod_{i<j}S(\theta_{i}-\theta_{j})F_{n}^{\mathcal{O}}(-\theta_{1},\dots,-\theta_{n})\]
The product appearing is nothing but the multi-particle R-matrix,
(\ref{Rmatrix}), which connects the \emph{in} and \emph{out} states.

Finally we use the fact that the reflection matrix can be considered
as a special form factor (of the identity operator $Id$) whose analytic
properties are well known. By definition\[
F_{2}^{Id}(\theta^{'}+i\pi,\theta)=\,_{out}\langle\theta^{'}\vert Id\vert\theta\rangle_{in}=R(\theta)\,_{out}\langle\theta^{'}\vert Id\vert-\theta\rangle_{out}=R(\theta)\delta(\theta+\theta^{'})\]
Now using the permutation property and tricks as above we have\[
F_{2}^{Id}(\theta^{'}+i\pi,\theta)=S(i\pi+\theta^{'}-\theta)F_{2}^{Id}(\theta,\theta^{'}+i\pi)=S(i\pi+\theta^{'}-\theta)R(i\pi+\theta^{'})\delta(\theta+\theta^{'})\]
which, due to the boundary crossing unitarity, is equivalent to the
previous expression.

\subsection{General solution}

In this section we describe the general procedure we use to obtain
the solutions of the form factor equations in the various specific
models. In doing so we emphasize the similarities and the differences
between the boundary and bulk form factors and also separate the (boundary)
operator dependent parts from the ones that depend on the specific
field theory considered but are independent of the operators in question.

\subsubsection{One particle form factors}

In sharp contrast to the bulk case, in the boundary theory, the boundary
operators in general may have non trivial one particle form factors
(1PFF). Since the multi-particle form factors are recursively determined,
the 1PFF-s are very important inputs to these recursions, and their
determination is necessarily the first step. The equations for the
1PFF read: \begin{equation}
F_{1}(\theta)=R(\theta)F_{1}(-\theta)\quad;\quad F_{1}(i\pi+\theta)=R(-\theta)F_{1}(i\pi-\theta),\label{eq:1pff}\end{equation}
 where the reflection amplitude $R(\theta)$ is analytic in the physical
strip $0\leq\Im m(\theta)\leq\pi/2$ (apart from the presence of finitely
many discrete poles on the imaginary axis), and from general considerations
using the reduction formulae we know that $F_{1}(\theta)$ is analytic
for $0\leq\Im m(\theta)\leq\pi$. Note that if $F_{1}(\theta)$ is
a solution of (\ref{eq:1pff}) then $F_{1}(\theta)\Psi(\theta)$ is
also a solution provided \[
\Psi(\theta)=\Psi(-\theta),\qquad\Psi(i\pi+\theta)=\Psi(i\pi-\theta),\]
 i.e. if $\Psi$ is even and $2\pi i$ periodic. Therefore one can
take $\Psi(\theta)=\psi(y)$ with $y=e^{\theta}+e^{-\theta}$.

To construct solutions to (\ref{eq:1pff}) we reduce them to a problem
already solved in the bulk form factor bootstrap. To this end we write
$F_{1}(\theta)=g_{1}(\theta)g_{2}(i\pi-\theta)$ and suppose that
\begin{equation}
g_{1}(\theta)=R(\theta)g_{1}(-\theta)\quad;\quad g_{1}(i\pi+\theta)=g_{1}(i\pi-\theta),\label{eq:g1}\end{equation}
 which are nothing else but the bulk two particle form factor equations
\cite{FMS}, where the reflection amplitude, $R(\theta)$, plays the
role of the S-matrix. Furthermore, plugging this product form $F_{1}$
into (\ref{eq:1pff}) reveals, that $g_{2}$ must also solve (\ref{eq:g1}).
Thus a solution to (\ref{eq:1pff}) can be constructed as \[
F_{1}(\theta)=g(\theta)g(i\pi-\theta),\]
 where $g(\theta)$ is an appropriate solution of (\ref{eq:g1}).

To obtain a solution of (\ref{eq:g1}) we use the following theorem
\cite{KW}. If the function $h(\theta)$ is meromorphic in the physical
strip $0\leq\Im m(\theta)<\pi$ with possible poles at $i\alpha_{1},\dots,i\alpha_{l}$
and zeros at $i\beta_{1},\dots,i\beta_{k}$ and grows as at most a
polynomial in $\exp(|\theta|)$ for $|\Re e\,\theta|\rightarrow\infty$,
furthermore it satisfies \[
h(\theta)=M(\theta)h(-\theta);\quad M(\theta)=\exp\left\{ \int_{0}^{\infty}dtf(t)\sinh\left(\frac{t\theta}{i\pi}\right)\right\} ;\quad h(i\pi-\theta)=h(i\pi+\theta);\]
 then it is uniquely defined up to normalization as \[
h(\theta)=\frac{\prod_{j=1}^{k}\sinh\left(\frac{1}{2}(\theta-i\beta_{j})\right)\sinh\left(\frac{1}{2}(\theta+i\beta_{j})\right)}{\prod_{j=1}^{l}\sinh\left(\frac{1}{2}(\theta-i\alpha_{j})\right)\sinh\left(\frac{1}{2}(\theta+i\alpha_{j})\right)}\exp\left\{ \int_{0}^{\infty}dtf(t)\frac{\sin^{2}\left(\frac{i\pi-\theta}{2\pi}t\right)}{\sinh x}\right\} .\]
 Since the reflection amplitudes are usually expressed as products
of the blocks $(x_{i})$ \begin{equation}
R(\theta)=\prod_{i}(x_{i});\qquad(x_{i})=\frac{\sinh(\frac{\theta}{2}+i\frac{\pi x_{i}}{2})}{\sinh(\frac{\theta}{2}-i\frac{\pi x_{i}}{2})},\label{eq:blockdef}\end{equation}
 to use this theorem we need the integral representation of one single
block $(x)$,: \[
-(x)=\exp\left\{ 2\int_{0}^{\infty}\frac{dt}{t}\frac{\sinh t(1-x)}{\sinh t}\sinh\left(\frac{t\theta}{i\pi}\right)\right\} .\]
 Then, if $R(\theta)$ consists of an even number of blocks, the \emph{minimal}
solution (with \textsl{no zeroes and poles}) to eq.(\ref{eq:1pff})
can be written as \[
r_{\textrm{min}}^{e}(\theta)=\exp\left\{ 2\int_{0}^{\infty}\frac{dt}{t}\frac{\sum_{i}\sinh t(1-x_{i})}{\sinh^{2}t}\left(1-\cosh\frac{t}{2}\cos\frac{t}{\pi}\left(\frac{i\pi}{2}-\theta\right)\right)\right\} .\]
 If $R(\theta)$ contains an extra minus sign, or is the product of
an odd number of blocks, $R=-\prod(-(x_{i}))$, then the factor $g(\theta)$
necessarily contains a zero at the origin which is implemented by
putting an extra $\sinh\frac{\theta}{2}$ into it; thus in this case
$r_{\textrm{min}}^{o}(\theta)=\sinh\theta\, r_{\textrm{min}}^{e}(\theta)$.
In the following an important role is played by the appropriate modification
of the \textsl{\emph{minimal}} 1PFF denoted by $r(\theta)$ \[
r(\theta)=r_{\textrm{min}}(\theta)\times\frac{\mathrm{zeroes}}{\mathrm{poles}}\]
 where the last factor denotes an appropriate number of zeroes and
poles at the right places (usually the same as in $R(\theta)$).

Thus the general solution of eq.(\ref{eq:1pff}) can be written as
\[
F_{1}(\theta)=r(\theta)Q_{1}(y),\qquad y=e^{\theta}+e^{-\theta},\]
 where the choice of $Q_{1}(y)$ is restricted by the analyticity
and the possible asymptotics of $F_{1}$. It is the $Q_{1}(y)$ in
the 1PFF that carries the dependence on the boundary operator $\mathcal{O}$.
Note in particular that if $Q_{1}(y)$ corresponds to the operator
$\mathcal{O}$ then $\tilde{Q}_{1}(y)\sim y^{N}Q_{1}(y)$ with $N$
integer $N\geq1$, describes the 1PFF of the operator $\partial_{\tau}^{N}\mathcal{O}$.

\subsubsection{Two-particle form factors}

The next step is to investigate the two-particle form factors (2PFF).
The novel feature compared to the 1PFF is that their equations contain
also the bulk S-matrix. It is worthwhile to go through the analysis
in some detail since it is straightforward to write down the general
form of the $n$-particle form factors once that of the 2PFF-s is
obtained. The equations for the 2PFF-s have the form \begin{eqnarray}
F_{2}(\theta_{1},\theta_{2}) & = & S(\theta_{1}-\theta_{2})F_{2}(\theta_{2},\theta_{1}),\quad(a)\qquad F_{2}(\theta_{1},\theta_{2})=R(\theta_{2})F_{2}(\theta_{1},-\theta_{2})\quad(b)\nonumber \\
F_{2}(i\pi+\theta_{1},\theta_{2}) & = & R(-\theta_{1})F_{2}(i\pi-\theta_{1},\theta_{2}).\quad(c)\label{eq:2pff}\end{eqnarray}
 Note that if $F_{2}(\theta_{1},\theta_{2})$ is a solution to these
equations then so is $F_{2}(\theta_{1},\theta_{2})H(\theta_{1},\theta_{2})$
provided $H$ is a symmetric, even and $2i\pi$ periodic function.

To construct solutions to eq.(\ref{eq:2pff}) we write \[
F_{2}(\theta_{1},\theta_{2})=f(\theta_{1}-\theta_{2})\Psi(\theta_{1},\theta_{2})\]
 where $f(\theta)$ is the minimal bulk two particle form factor \cite{FMS},
i.e. the minimal solution of \[
f(\theta)=S(\theta)f(-\theta),\qquad f(i\pi+\theta)=f(i\pi-\theta).\]
 Plugging this $F_{2}$ into (\ref{eq:2pff}a) reveals that $\Psi$
must be symmetric $\Psi(\theta_{1},\theta_{2})=\Psi(\theta_{2},\theta_{1})$.
The most convenient way to satisfy (\ref{eq:2pff}b) is that $\Psi$
has the form \[
\Psi(\theta_{1},\theta_{2})=f(\theta_{1}+\theta_{2})r(\theta_{1})r(\theta_{2})\Phi(\theta_{1},\theta_{2})\]
 where $\Phi$ is symmetric and even $\Phi(\theta_{1},\theta_{2})=\Phi(\theta_{1},-\theta_{2})$.
Finally this $F_{2}$ satisfies eq.(\ref{eq:2pff}c) also if $\Phi(i\pi-\theta_{1},\theta_{2})=\Phi(i\pi+\theta_{1},\theta_{2})$.
The conditions on $\Phi$ can be satisfied simply by writing $\Phi(\theta_{1},\theta_{2})=\phi(y_{1},y_{2})$
where $\phi$ is a symmetric function of the $y_{i}$-s ($y_{i}=e^{\theta_{i}}+e^{-\theta_{i}}$,
$i=1,2$). Thus the general form of the 2PFF, compatible with eq.(\ref{eq:2pff})
is \[
F_{2}(\theta_{1},\theta_{2})=r(\theta_{1})r(\theta_{2})f(\theta_{1}-\theta_{2})f(\theta_{1}+\theta_{2})\phi(y_{1},y_{2}),\qquad\phi(y_{1},y_{2})=\phi(y_{2},y_{1}).\]
 Different choices of the boundary operator $\mathcal{O}$ correspond
to different functions $\phi(y_{1},y_{2})$ in this expression.

\subsubsection{Multi-particle form factors}

From the explicit form of the 2PFF it is clear that the general form
of the multi-particle form factors can be written in the following
form: \begin{equation}
F_{n}(\theta_{1},\theta_{2},\dots,\theta_{n})=G_{n}(\theta_{1},\theta_{2},\dots,\theta_{n})\prod_{i=1}^{n}r(\theta_{i})\prod_{i<j}f(\theta_{i}-\theta_{j})f(\theta_{i}+\theta_{j}),\label{eq:GenAnsatz}\end{equation}
 where $f(\theta)$ is the minimal bulk two particle form factor.
As a consequence of the form factor equations $G_{n}$ is a $2\pi i$
periodic, symmetric and even function of the rapidities: $\theta_{i}$,
i.e. it is symmetric in the variable $y_{i}=2\cosh\theta_{i}$. When
the bulk S-matrix is nontrivial, the bulk kinematical singularity
equations \begin{equation}
-i\mathop{\textrm{Res}}_{\theta^{'}=\theta}F_{n+2}(\theta^{'}+i\pi,\theta,\theta_{1},\dots,\theta_{n})=(1-\prod_{i=1}^{n}S(\theta-\theta_{i})S(\theta+\theta_{j}))F_{n}(\theta_{1},\dots,\theta_{n})\label{eq:ksing}\end{equation}
 give recursive relations linking $G_{n}$ to $G_{n+2}$. (Note that
these singularities are \textsl{absent} in the two particle case).
The advantage of using the $y_{i}$-s becomes clear if one tries to
describe the bulk kinematical singularities: since $y(i\pi+\theta)=-y(\theta)$,
thus including a (symmetric) factor $y_{i}+y_{j}$ in the denominator
automatically accounts for the pole. Therefore in the following we
put \[
G_{n}(\theta_{1},\theta_{2},\dots,\theta_{n})=\frac{Q_{n}(y_{1},y_{2}\dots,y_{n})}{\prod\limits _{i<j}(y_{i}+y_{j})},\]
 (with $Q_{n}$ being a symmetric function of $y_{1},\dots,y_{n}$)
and then eq.(\ref{eq:ksing}) give recursive relations between the
functions $Q_{n}$. Clearly the actual form of these recursive relations
varies from model to model since they depend on the bulk S-matrix.
The form of the recursions depends also on the choice of the 1PFF
$r(\theta)$; it is useful to choose an $r(\theta)$ which gives the
simplest possible recursion. Writing the 2PFF in the same form as
the $n$-particle one \[
\phi(y_{1},y_{2})=\frac{Q_{2}(y_{1},y_{2})}{y_{1}+y_{2}}\]
 then the absence of kinematical singularities requires $Q_{2}(y,-y)=0$. 

If the bulk S-matrix is nontrivial and the reflection factor has a
pole at $\frac{i\pi}{2}$ then the form factors with odd and even
particle number are connected by the boundary kinematical singularity
equation:\[
-i\mathop{\textrm{Res}}_{\theta=0}F_{n+1}^{\mathcal{O}}(\theta+\frac{i\pi}{2},\theta_{1},\dots,\theta_{n})=\frac{g}{2}\Bigl(1-\prod_{i=1}^{n}S\bigl(\frac{i\pi}{2}-\theta_{i}\bigr)\Bigr)F_{n}^{\mathcal{O}}(\theta_{1},\dots,\theta_{n})\]
The corresponding pole in the $n$ particle form factor can be included
as \[
G_{n}(\theta_{1},\theta_{2},\dots,\theta_{n})=\frac{Q_{n}(y_{1},y_{2}\dots,y_{n})}{\prod_{i}y_{i}\,\prod\limits _{i<j}(y_{i}+y_{j})},\]
and the boundary kinematical singularity equation relates $Q_{n}$
to $Q_{n+1}$.

The even and the odd particle form factors are also related if the
bulk S-matrix has a \lq\lq self fusing'' pole describing the 2
particle $\rightarrow$ 1 particle process, which parallels the bulk
situation (this happens e.g. in the Lee-Yang model). (In this case
it is customary to include this pole also in $f(\theta)$). Since
the fusing angle in this process is necessarily $2\pi/3$, one finds
from bootstrap that in this case the dynamical singularities imply
\begin{equation}
-i\mathop{\textrm{Res}}_{\theta^{'}=\theta}F_{n+2}(\theta^{'}+\frac{i\pi}{3},\theta-\frac{i\pi}{3},\theta_{1},\dots,\theta_{n})=\Gamma F_{n+1}(\theta,\theta_{1},\dots,\theta_{n}),\label{eq:dsing}\end{equation}
 where $\Gamma$ is related to the residue of the S-matrix at the
self fusing pole: $-i\textrm{res}_{\theta=\frac{2\pi i}{3}}S(\theta)=\Gamma^{2}$.

An important restriction follows on the form factor functions from
requiring a power law bounded ultraviolet behaviour for the two point
correlator of the boundary operators $\langle0|\mathcal{O}(\tau)\mathcal{O}(0)|0\rangle$:
the growth of the function $F_{n}(\theta_{1},\dots,\theta_{n})$ must
be bounded by some exponential of the rapidity as $\theta\rightarrow\infty$
(i.e. the form factors only grow polynomially with particle energy).
This can be shown using an argument identical to that in the bulk
case \cite{counting_ops}. If $r\left(\theta\right)$ and $f\left(\theta\right)$
are specified in a way to include all poles induced by the dynamics
of the model, then it follows that the functions $Q_{n}$ must be
\textsl{polynomials} of the $y_{i}$. Therefore in the following we
only look for explicit \textsl{polynomial} solutions of the various
recursion equations. This is a posteriori confirmed since we find
as many polynomial solution of the boundary form factor equation as
many independent local operator exist in the theories.

\subsection{Two-point function}

Once an appropriate solution of the form factor axioms is found it
can be used to describe correlators of boundary operators. The two-point
function of the boundary operator $\mathcal{O}$ can be computed by
inserting a complete set of states \begin{equation}
\langle0\vert\mathcal{O}(t)\mathcal{O}(0)\vert0\rangle=\sum_{n=0}^{\infty}\frac{1}{(2\pi)^{n}}\int_{\theta_{1}>\theta_{2}>\dots>\theta_{n}>0}d\theta_{1}d\theta_{2}\dots d\theta_{n}e^{-imt\sum_{i}\cosh\theta_{i}}F_{n}F_{n}^{+}\label{eq:2pt}\end{equation}
where time translation invariance was used and the form factors are
\[
F_{n}=\langle0\vert\mathcal{O}(0)\vert\theta_{1},\theta_{2},\dots,\theta_{n}\rangle_{in}=F_{n}^{\mathcal{O}}(\theta_{1},\theta_{2},\dots,\theta_{n})\]
and \[
F_{n}^{+}=\,_{in}\langle\theta_{1},\theta_{2},\dots,\theta_{n}\vert\mathcal{O}(0)\vert0\rangle=F_{n}^{\mathcal{O}}(i\pi+\theta_{n},i\pi+\theta_{n-1},\dots,i\pi+\theta_{1})\]
 which, for unitary theories, is the complex conjugate of the previous
one: $F_{n}^{+}=F_{n}^{*}$. In the Euclidean $(\tau=it)$ version
of the theories the form factor expansion of the correlator for large
separations converges rapidly since multi-particle terms are exponentially
suppressed. 

The identification between solutions of the form factor equations
and operators of the theory is a central issue. One possible way is
to analyze the behaviour of the boundary correlators for short distances.
Although on general grounds one may expect the form factor expansion
to converge rapidly only in the infrared (large volume) regime, the
examples from the various bulk theories, where the form factor expansion
converges even in the UV domain (see e.g. \cite{Z1}), suggest that
similar behaviour may happen in the boundary setting as well. If the
theory can be described as a relevant perturbation of a conformal
field theory, then in the UV domain the two-point function must follow
a behaviour dictated by this limiting theory. The short distance singularity
exponent is related to the scaling dimension of the operator $\mathcal{O}$
and can be calculated from the asymptotic growth of the form factors.

\section{Model studies}

In this section we carry out a detailed investigation of the solutions
of the form factor equations in four different models. The first two
models (the massive scalar field with linear boundary interaction
and the Ising model interacting with a boundary magnetic field \cite{GZ})
are free in the bulk and the correlation functions are known explicitly,
thus the form factors obtained from the explicit field theoretic solutions
can be compared directly to the solutions of the form factor equations.
In both cases we find that the space of appropriate \textsl{polynomial}
solutions of the FF equations can be identified with the space of
local boundary operators obtained from the explicit construction.
In the case of the Ising model we also show how the conformal dimensions
of the various operators of the ultraviolet limiting BCFT can be obtained
from the solutions of the FF equations. The third and fourth models,
namely the scaling Lee-Yang and the sinh-Gordon models with integrability
preserving boundaries are among the simplest boundary integrable theories.
In contrast to the previous cases they cannot be solved directly so
one has to rely upon the solution of the form factor equations. Since
these models contain nontrivial bulk interactions the recursion relations
connecting the multi-particle form factors are no longer trivial,
and in these cases we investigate their solutions in detail.

\subsection{Massive scalar with linear boundary interaction}

\subsubsection{Direct calculation}

The free massive scalar field $\Phi(x,t)$ restricted to the negative
half-line $x\leq0$ subject to linear boundary condition\begin{equation}
\partial_{x}\Phi(x,t)|_{x=0}=-\lambda(\Phi(0,t)-\Phi_{0}).\label{eq:robin}\end{equation}
 can be described by the following Lagrangian:\[
\mathcal{L}=\Theta(-x)\left(\frac{1}{2}(\partial_{t}\Phi)^{2}-\frac{1}{2}(\partial_{x}\Phi)^{2}-\frac{m^{2}}{2}\Phi^{2}\right)-\delta(x)\frac{\lambda}{2}(\Phi-\Phi_{0})^{2},\]
This one parameter family of linear boundary condition interpolates
between Neumann $\partial_{x}\Phi\vert_{x=0}=0$ (for $\lambda=0$)
and Dirichlet $\Phi\vert_{x=0}=\Phi_{0}$ (for $\lambda\to\infty$)
boundary conditions. Since for any $\lambda$ we are dealing with
a free theory it can be solved explicitly. The mode decomposition
of the field is \[
\Phi(x,t)=Ae^{mx}+\int_{0}^{\infty}\frac{dk}{\omega(k)}\Bigl\{ a(k)e^{-i\omega(k)t}\bigl(e^{ikx}+R(k)e^{-ikx}\bigr)+a^{+}(k)e^{i\omega(k)t}\bigl(e^{-ikx}+R(-k)e^{ikx}\bigr)\Bigr\}\]
 where $A=\frac{\lambda}{m+\lambda}\Phi_{0}$ and \[
R(k)=\frac{k-i\lambda}{k+i\lambda}\]
 is the reflection factor on the boundary at $x=0$. The creation/annihilation
operators are normalized as \[
[a(k),a^{+}(k^{'})]=2\pi\omega(k)\delta(k-k^{'})\quad,\quad k\,,\: k^{'}>0\]
The boundary two-point function can be calculated easily\[
\langle0|\Phi(0,t)\Phi(0,t^{'})|0\rangle=A^{2}+\int_{0}^{\infty}\frac{dk}{2\pi\omega(k)}e^{-i\omega(k)(t-t^{'})}\left(1+R(k)\right)\left(1+R(-k)\right)\]
 By comparing this expression to the form factor expansion of the
two-point function (\ref{eq:2pt}), the form factor of the elementary
field can be extracted: \[
\langle0|\Phi(0,t)|\theta\rangle=e^{-i\omega(k)t}\left(1+R(k)\right)\]
The same result can be obtained by taking the general (space-dependent)
two point function \begin{eqnarray*}
\langle0|T\left(\Phi(x,t)\Phi(x^{'},t^{'})\right)|0\rangle & = & A^{2}e^{m(x+x^{'})}\\
 &  & +\int\frac{d^{2}k}{(2\pi)^{2}}\frac{i}{k^{2}-m^{2}+i\epsilon}e^{-ik_{0}(t-t^{'})}\left(e^{-ik_{1}(x-x^{'})}+R(k)e^{+ik_{1}(x+x^{'})}\right)\end{eqnarray*}
 and using the boundary reduction formula \cite{BBT} \begin{eqnarray*}
\langle0|\Phi(x,t)|\theta\rangle & = & 2i\int_{-\infty}^{0}dx^{'}\int_{-\infty}^{\infty}dt^{'}\: e^{i\omega(\theta)t^{'}}\cos(p(\theta)x^{'})\left\{ \partial_{t^{'}}^{2}-\partial_{x^{'}}^{2}+m^{2}+\delta(x^{'})\partial_{x^{'}}\right\} \\
 &  & \hspace{8cm}\langle0|T\left(\Phi(x,t)\Phi(x^{'},t^{'})\right)|0\rangle\end{eqnarray*}
 where $\omega(\theta)=m\cosh(\theta)$ and $p(\theta)=m\sinh(\theta)$.
Performing explicitly the calculation yields\begin{equation}
\langle0|\Phi(x,t)|\theta\rangle=e^{-i\omega(\theta)t}(e^{ip(\theta)x}+R(\theta)e^{-ip(\theta)x})\label{eq:Phivev}\end{equation}
which for the form factor of the operator $\Phi(x=0,t)$ reads as 

\[
\langle0|\Phi(0,t)|\theta\rangle=e^{-i\omega(\theta)t}\left(1+R(\theta)\right)\]
 Introducing $\tau=-it$ one also gets\[
\langle0|\partial_{\tau}^{n}\Phi(0,0)|\theta\rangle=\omega(\theta)^{n}\left(1+R(\theta)\right),\quad n>0.\]
 Clearly these operators have no multi-particle matrix elements. It
is important to realize that $\partial_{x}\Phi(0,0)$ is not an independent
operator since the boundary condition eq.(\ref{eq:robin}) relates
it to $\Phi(0,0)$, thus the set of independent boundary operators
having only one particle matrix elements is given by $\partial_{\tau}^{n}\Phi(0,0)$.
To obtain multi-particle matrix elements one has to consider \[
\langle0|:\Phi(x_{1},t_{1})\dots\Phi(x_{k},t_{k}):|\theta_{1}\dots\theta_{k}\rangle.\]
 Using the analogous boundary reduction formula and the Wick theorem
we obtain \begin{eqnarray}
 &  & \langle0|:\Phi(x_{1},t_{1})\dots\Phi(x_{k},t_{k}):|\theta_{1}\dots\theta_{k}\rangle=\label{eq:Phisvev}\\
 &  & \left\{ e^{-i\omega(\theta_{1})t_{1}}\left(e^{ip(\theta_{1})x_{1}}+R(\theta_{1})e^{-ip(\theta_{1})x_{1}}\right)\right\} \dots\left\{ e^{-i\omega(\theta_{k})t_{k}}\left(e^{ip(\theta_{k})x_{k}}+R(\theta_{k})e^{-ip(\theta_{k})x_{k}}\right)\right\} +\dots,\nonumber \end{eqnarray}
 where the ellipses at the end represent additional terms which make
it completely symmetric in all coordinates. From this expression one
can extract the form factor of the most general boundary operator
of the theory\[
\langle0|:\partial_{\tau_{1}}^{n_{1}}\Phi(0,0)\dots\partial_{\tau_{k}}^{n_{k}}\Phi(0,0):|\theta_{1}\dots\theta_{k}\rangle=\omega(\theta_{1})^{n_{1}}\left(1+R(\theta_{1})\right)\dots\omega(\theta_{k})^{n_{k}}\left(1+R(\theta_{k})\right)+\dots\]
 where again a complete symmetrization in the $\theta_{i}$ rapidities
is understood. Checking the leading asymptotic behaviour of these
form factors gives that for all $\theta_{i}\sim\theta$ large they
grow as $e^{N\theta}$, where $N=n_{1}+\dots+n_{k}$ is the total
number of derivatives in the expression. We note that we have as many
operators for a given $N$ as many partition $N$ has into the numbers
$1,2,\dots,k$. This can be seen by writing $N=N_{1}+2N_{2}+\dots+kN_{k}$
and associating to it the operator with $n_{1}=N_{1}+N_{2}\dots+N_{k}$,
$n_{2}=N_{2}+\dots+N_{k}$ $\dots$ $n_{k}=N_{k}$ derivatives.

The Dirichlet boundary condition ($R=-1$) can be obtained in the
$\lambda\to\infty$ limit. Clearly $\Phi\vert_{x=0}=\Phi_{0}$ is
a c-number and the Dirichlet boundary condition does not connect the
operator $\partial_{x}\Phi\vert_{x=0}$ to $\Phi\vert_{x=0}$. We
can extract, however, the form factors of the operator $\partial_{x}\Phi(0,t)$
from that of $\Phi(0,t)$ by taking the $\lambda\to\infty$ limit
carefully in (\ref{eq:Phivev}): \[
\langle0|\partial_{x}\Phi(0,t)|\theta\rangle=e^{-i\omega(\theta)t}2ip(\theta)\]
and for its derivatives \[
\langle0|\partial_{\tau}^{n}\partial_{x}\Phi(0,0)|\theta\rangle=\omega(\theta)^{n}2ip(\theta),\quad n>0.\]
 This can be extended similarly to the most general operator as \[
\langle0|:\partial_{\tau}^{n_{1}}\partial_{x}\Phi(0,0)\dots\partial_{\tau}^{n_{k}}\partial_{x}\Phi(0,0):|\theta_{1}\dots\theta_{k}\rangle=\omega(\theta_{1})^{n_{1}}2ip(\theta_{1})\dots\omega(\theta_{k})^{n_{k}}2ip(\theta_{k})+\dots\]
 where again a complete symmetrization in the $\theta_{i}$ rapidities
is understood. Checking the leading asymptotic behaviour of these
form factors gives that for all $\theta_{i}\sim\theta$ large they
grow as $e^{N\theta}$, where $N=k+n_{1}+\dots+n_{k}$ is the total
number of derivatives in the expression.

\subsubsection{Solving the form factor equations}

The bulk $S$-matrix of the theory together with the reflection factor
are \[
S(\theta)=1,\quad R(\theta)=\frac{\sinh\theta-i\frac{\lambda}{m}}{\sinh\theta+i\frac{\lambda}{m}}=-\left(1+\frac{B}{2}\right)\left(-\frac{B}{2}\right),\quad\frac{\lambda}{m}=\sin\frac{\pi B}{2},\]
where the block notation (\ref{eq:blockdef}) is used to express $R(\theta)$.
As a consequence of $S=1$ the minimal bulk form factor is trivial:
$f(\theta)=1$. To determine the 1PFF we note that this reflection
factor is identical to the two particle S-matrix of the sinh-Gordon
model if the above identification of parameters is done. Therefore
eq.(\ref{eq:g1}) in this case is identical to the equation for the
minimal bulk form factor of the sinh-Gordon model. Choosing for $g(\theta)$
the solution given in \cite{FMS} (described in detail in the sinh-Gordon
section) gives \[
r(\theta)=2g(\theta)g(i\pi-\theta)=\frac{2\sinh\theta}{\sinh\theta+i\frac{\lambda}{m}}=1+R(\theta)\]
 Clearly this corresponds to the form factor of the operator $\Phi(0,0)-A$. 

Now we demonstrate that the number of independent solutions of the
form factor equations coincides with the number of local boundary
operators. In this case the general Ansatz (\ref{eq:GenAnsatz}) takes
the following form\[
F_{n}(\theta_{1},\dots,\theta_{n})=P_{n}(y_{1},\dots,y_{n})\prod_{i}r(\theta_{i})\]
 where $y_{i}=2\cosh\theta_{i}$ as before. Since the bulk $S$-matrix
is trivial there are no bulk/boundary kinematical singularities and
$P_{n}$ is a completely symmetric polynomial in the $y_{i}$-s. One
can count the independent solutions of the BFF equations by counting
the possible solutions for $P_{n}$. If $P_{n}$ has degree $N$ then
the solutions are given by the partitions of $N$ into the numbers
$1,2,\dots n$ in the following way. Since the completely symmetric
polynomials of $n$ variable are generated by the $\sigma_{i}$-s
(elementary symmetric polynomials of degree $i$) one can write:\[
\prod_{i=1}^{n}(x+x_{i})=\sum_{i=1}^{n}\sigma_{i}\, x^{n-i};\quad P_{n}\propto\sigma_{1}^{k_{1}}\sigma_{2}^{k_{2}}\dots\sigma_{n}^{k_{n}};\quad N=\sum ik_{i}\]
 It is clear that this space has the same dimension as the space of
boundary operators having only $n$ particle matrix elements with
asymptotic growth $e^{N\theta}$ . 

The Dirichlet ($\lambda\to\infty$) limit for the operator $\partial_{x}\Phi(0,0)$
can be obtained using its relation to $\Phi(0,0)$ via the boundary
condition (\ref{eq:robin}) as we did in the Lagrangian framework.

\subsection{Ising model with boundary magnetic field }

\subsubsection{Direct calculation}

The Ising model with a boundary magnetic field can be described by
a free massive Majorana fermion perturbed at the boundary \cite{GZ}.
In Minkowskian formalism the Dirac equation can be obtained form the
Lagrangian:\[
\mathcal{L}=\frac{1}{2}\bar{\Psi}(i\gamma^{\mu}\partial_{\mu}-m)\Psi\]
The gamma matrices are chosen as \[
\gamma^{0}=\left(\begin{array}{cc}
0 & -i\\
i & 0\end{array}\right)\quad;\quad\gamma^{1}=\left(\begin{array}{cc}
0 & -i\\
-i & 0\end{array}\right)\]
in order for the Dirac equation to be real:\[
\left(\begin{array}{cc}
-m & \partial_{x}+\partial_{t}\\
\partial_{x}-\partial_{t} & -m\end{array}\right)\Psi=0\]
Thus the Majorana condition corresponds to taking $\Psi$ real. Using
the component notation \[
\Psi=\left(\begin{array}{c}
\psi_{+}\\
\psi_{-}\end{array}\right)\]
the Lagrangian of the boundary field theory takes the form 

\[
\mathcal{L}=-i\Theta(-x)\left(\frac{1}{2}\psi_{+}(\partial_{t}-\partial_{x})\psi_{+}-\frac{1}{2}\psi_{-}(\partial_{t}+\partial_{x})\psi_{-}-m\psi_{+}\psi_{-}\right)-i\delta(x)U_{B}\]
where \[
U_{B}=\frac{1}{2}\psi_{+}\psi_{-}+\frac{1}{2}a\dot{a}+\frac{1}{2}ha(\psi_{+}+\psi_{-})\]
The operator $a$ is a boundary fermion $a^{2}=1$, which implements
the boundary condition \[
\partial_{t}(\psi_{+}-\psi_{-})=\frac{h^{2}}{2}(\psi_{+}+\psi_{-})\]
Since the theory is free we can solve it explicitly. The mode expansion
of the fermionic fields are \begin{eqnarray*}
\psi_{\pm}(x,t) & = & \int_{0}^{\infty}\frac{d\theta}{2\pi}\Biggl\{ b(\theta)e^{-i\omega(\theta)t}\biggl(u_{\pm}(\theta)e^{ik(\theta)x}+R(\theta)u_{\pm}(-\theta)e^{-ik(\theta)x}\biggr)\\
 &  & \hspace{2cm}+b^{+}(\theta)e^{i\omega(\theta)t}\biggl(v_{\pm}(\theta)e^{-ik(\theta)x}+R(-\theta)v_{\pm}(-\theta)e^{ik(\theta)x}\biggr)\Biggr\}\end{eqnarray*}
where $u_{\pm}(\theta)=v_{\pm}^{*}(\theta)=\sqrt{m}e^{\mp\frac{i\pi+2\theta}{4}}$
are the spinor amplitudes, $R(\theta)$ is nothing else but the one-particle
reflection factor \[
R(\theta)=i\tanh\left(\frac{i\pi}{4}-\frac{\theta}{2}\right)\frac{\sinh\theta+i\kappa}{\sinh\theta-i\kappa}\qquad,\quad\kappa=1-\frac{h^{2}}{2m}\]
 and the creation/annihilation operators are normalized as \[
\{ b(\theta),b^{+}(\theta^{'})\}=2\pi\delta(\theta-\theta^{'})\]
The boundary two point function can be calculated explicitly: \begin{equation}
\langle0|\psi_{\pm}(0,t)\psi_{+}(0,t^{'})|0\rangle=\int_{0}^{\infty}\frac{d\theta}{2\pi}e^{-i\omega(\theta)(t-t^{'})}\left(u_{\pm}(\theta)+R(\theta)u_{\pm}(-\theta)\right)\left(v_{+}(\theta)+R(-\theta)v_{+}(-\theta)\right)\label{eq:Fermcorrexact}\end{equation}
which, as compared to the form factor expansion (\ref{eq:2pt}), gives
\begin{equation}
\langle0\vert\psi_{\pm}(0,t)\vert\theta\rangle=e^{-i\omega(\theta)t}\left(u_{\pm}(\theta)+R(\theta)u_{\pm}(-\theta)\right)\label{eq:Fermionffexact}\end{equation}
These two operators are not independent since they are related by
the boundary condition and so there is only one boundary fermion field,
say $\psi_{+}$. As a result, the algebraically independent operators
at the boundary are the fermion field and its derivatives $\partial_{\tau}^{n}\psi_{+}$.
Note that $\partial_{x}\psi_{+}|_{x=0}$ is not an independent field,
as it is determined by the Dirac equation in terms of $\partial_{\tau}\psi_{+}|_{x=0}$
and $\psi_{-}|_{x=0}$. As a consequence of the fermionic nature of
the field the most general boundary operator has the form $\partial_{\tau}^{n_{1}}\psi_{+}\partial_{\tau}^{n_{2}}\psi_{+}\dots\partial_{\tau}^{n_{k}}\psi_{+}$
where $n_{1}>n_{2}>\dots>n_{k}$ (the inequalities are strict, in
contrast to the bosonic case discussed earlier). $N=n_{1}+n_{2}+\dots+n_{k}$
is called the level of the operator, and operators at level $N$ can
be brought in one-to-one correspondence with partitions of $N$ into
the numbers $1,2,\dots,k$. For a partition \[
N=kN_{k}+(k-1)N_{k-1}+\dots+2N_{2}+N_{1}\]
we associate the operator above with $n_{k}=N_{k};\, n_{k-1}=N_{k}+N_{k-1};\dots;\, n_{1}=N_{k}+N_{k-1}+\dots+N_{1}$. 

(Had we included also the operator $a$ we would have had to perform
a GSO type projection, leaving only non-fermionic operators. This
would amount to keeping all operators with an even number of fermion
factors plus all odd ones multiplied with a factor $a$, but this
would lead to the same number of operators.)

\subsubsection{Solution of the FF bootstrap}

Using again the block notation (\ref{eq:blockdef}) the $S$-matrix
of the theory and the reflection factor are \cite{GZ}\[
S(\theta)=-1\quad,\qquad R_{x}(\theta)=[x]\left(-\frac{1}{2}\right)\quad,\qquad[x]=(x)(1-x)\]
 where $x$ is related to the boundary magnetic field as \[
\sin\pi x=1-\frac{h^{2}}{2m}=\kappa\]
 For $h=0$ we recover the free boundary condition with \[
R_{\textrm{free}}(\theta)=\left(\frac{1}{2}\right)\]
which has a pole at $i\frac{\pi}{2}$ corresponding to the fact that
the ground state is doubly degenerate. In contrast to the generic
situation the pole at $i\frac{\pi}{2}$ is a \emph{dynamical} pole
and not a kinematical one (since the field has no vacuum expectation
value at all). The $h\rightarrow\infty$ limit corresponds to the
fixed boundary condition (when the Ising spin takes a fixed value
at the boundary), and the reflection factor is \[
R_{\textrm{fixed}}(\theta)=\left(-\frac{1}{2}\right)\]
which has no pole in the physical strip at all. The minimal one particle
form factor for the fixed case can be calculated directly using the
recipe in Section 2 \[
r_{\textrm{fixed}}(\theta)=\frac{\sinh\theta}{\sinh(\frac{\theta}{2}+i\frac{\pi}{4})}\]
For the free case we include the dynamical singularity into the 1PFF
\[
r_{\textrm{free}}(\theta)=-2i\frac{\sinh\theta}{\cosh\theta}\sinh\left(\frac{\theta}{2}+i\frac{\pi}{4}\right)\]
 The simplest solution which interpolates between these two cases
and has a pole exactly at the location of the boundary dynamical singularity
of the reflection factor is \begin{equation}
r(\theta)=\frac{\sinh\theta}{\sinh(\frac{\theta}{2}+i\frac{\pi}{4})}\frac{\cosh\theta+i(1-\kappa)}{\sinh\theta-i\kappa}.\label{eq:rising}\end{equation}
 This expression is the same we obtained from the exact solution of
the model (\ref{eq:Fermionffexact}). The minimal bulk two particle
form factor is simply \[
f(\theta)=\sinh\frac{\theta}{2}.\]
 Since $R(\theta)$ has no kinematical pole at $i\frac{\pi}{2}$,
boundary kinematical singularities are absent, and since the bulk
$S$-matrix is $-1$ there are no bulk kinematical singularities either.
Thus we look for the $n$ particle form factors in the form (\ref{eq:GenAnsatz})
\[
F_{n}(\theta_{1},\dots,\theta_{n})=P_{n}(y_{1},\dots,y_{n})\prod_{i}r(\theta_{i})\prod_{i<j}f(\theta_{i}-\theta_{j})f(\theta_{i}+\theta_{j}),\]
 where $y_{i}=2\cosh\theta_{i}$ and $P_{n}$ is a completely symmetric
polynomial in the $y_{i}$-s. Taking into account the special form
of $f(\theta)$ the form factor simplifies to \begin{equation}
F_{n}(\theta_{1},\dots,\theta_{n})=P_{n}(y_{1},\dots,y_{n})\prod_{i}r(\theta_{i})\prod_{i<j}(y_{i}-y_{j})\label{eq:isingff}\end{equation}
 The independent solutions are counted in the same way as in the bosonic
case, i.e. by the partitions of $N$ into the numbers $1,2,\dots,n$
and are generated by the $\sigma_{i}$-s. It is clear that the dimension
of the space they span is the same as the one of the boundary operators
obtained from the direct calculation. 

Since the UV limit of this theory is a boundary conformal field theory
one can go further than in the bosonic case and calculate the UV dimension
of the various boundary operators. As the form factor equations are
not coupled we can choose a basis among operators consisting of those
having matrix elements only with a certain fixed number of particles.

If the operator has only one-particle matrix element then its correlator
is \begin{equation}
\langle0|\mathcal{O}(\tau)\mathcal{O}(0)|0\rangle=\int_{0}^{\infty}\frac{d\theta}{2\pi}|F_{1}^{\mathcal{O}}(\theta)|^{2}e^{-m\cosh\theta\tau}\label{eq:isingtpf}\end{equation}
 where $F_{1}^{\mathcal{O}}(\theta)=r(\theta)P_{1}(y)$. Plugging
(\ref{eq:rising}) into (\ref{eq:isingtpf}) we obtain the exact correlator
(\ref{eq:Fermcorrexact}). If the operator $\mathcal{O}$ goes to
a scaling operator in the UV limit ($\tau\rightarrow0$) then the
exact correlator has the short distance asymptotics $\tau^{-2\Delta}$,
where $\Delta$ is the appropriate scaling dimension in the ultraviolet
BCFT. In (\ref{eq:isingtpf}) the singularity comes from the large
$\theta$ asymptotics of the form factor. If $|F_{1}^{\mathcal{O}}(\theta)|^{2}$
diverges as $y^{n}$ here, then the corresponding weight is $\Delta=\frac{n}{2}$.
Taking the simplest solution $P_{1}(y)=1$ the weight is $\Delta=\frac{1}{2}$
which corresponds to the boundary fermion field. Choosing $P_{1}(y)=\sigma_{1}^{n}(y)$
corresponds to the $n$-th derivative of this operator which has weight
$n+\frac{1}{2}$.

Similarly we can analyze an operator having $n$-particle matrix element
only. The corresponding correlator is \[
\langle0|\mathcal{O}_{(n)}(\tau)\mathcal{O}_{(n)}(0)|0\rangle=\int_{0}^{\infty}\frac{d\theta_{1}}{2\pi}\dots\int_{0}^{\infty}\frac{d\theta_{n}}{2\pi}\frac{1}{n!}|F_{n}^{\mathcal{O}}(\theta_{1},\dots,\theta_{n})|^{2}e^{-m(\cosh\theta_{1}+\dots+\cosh\theta_{n})\tau}\]
 The operator which has the mildest UV behaviour corresponds to $P_{n}=1$.
The corresponding form factor square for large $\theta$-s behaves
as \[
|F_{n}^{\mathcal{O}}(\theta_{1},\dots,\theta_{n})|^{2}\propto\mathrm{exp}(\theta(n+n(n-1))=e^{\theta n^{2}},\]
 thus the UV dimension of $\mathcal{O}_{(n)}$ is $\Delta=\frac{n^{2}}{2}$.
The explicit boundary operator which has nonzero matrix elements only
with $n$ particle states and has the mildest UV behaviour is \[
\psi_{+}\partial_{\tau}\psi_{+}\dots\partial_{\tau}^{n-1}\psi_{+}\]
with dimension $\Delta=\frac{n}{2}+\frac{n(n-1)}{2}$; therefore it
can be associated to $\mathcal{O}_{(n)}$.

To match the descendent operators, note that to any partition of $N=k_{1}+2k_{2}+\dots+n\cdot k_{n}$
there exists a solution of the form factor equations with $P_{n}^{N}=\sigma_{1}^{k_{1}}\dots\sigma_{n}^{k_{n}}$.
The number of such polynomials is the same as the number of descendants
of $\mathcal{O}_{(n)}$ at level $N$: to the given partition we can
associate the operator \[
\partial_{\tau}^{k_{n}}\psi_{+}\partial_{\tau}^{1+k_{n-1}+k_{n}}\psi_{+}\dots\partial_{\tau}^{n-1+k_{1}+\dots+k_{n}}\psi_{+}\]
Conversely, given an operator of the form \[
\partial_{\tau}^{p_{1}}\psi_{+}\dots\partial_{\tau}^{p_{n}}\psi_{+}\qquad,\qquad0\leq p_{1}<p_{2}\dots<p_{n}\]
of weight $N+\frac{n^{2}}{2}$, one can define a partition as $k_{n}=p_{1}$,
$k_{n-1}=p_{2}-k_{n}-1$, $\dots$ and thus associate a polynomial
solution of the form factor equations with appropriate asymptotic
behaviour. It is important to emphasize that we do not claim that
the form factor related to $P_{n}^{N}$ belongs to the operator above,
what we have shown is only that the dimension of the space of operators
with certain scaling dimension is the same as the dimension of the
solution of the form factor equations.

\subsection{The boundary scaling Lee-Yang model}

The scaling Lee-Yang model with boundary is a combined bulk and boundary
perturbation of the boundary version of the ${\mathcal{M}}(2/5)$
minimal model \cite{DPTW1pt,DPTW1}. In the bulk the perturbation
is given by the unique relevant spinless field $\phi$, at the boundary
the perturbation depends on which of the two possible conformal boundary
conditions is present in the unperturbed model. One, denoted by $\textrm{1}$
in \cite{DPTW1}, does not have any relevant boundary fields - thus
can have no boundary perturbation either -, while the other, denoted
$\Phi$ in \cite{DPTW1}, has a single relevant boundary field $\varphi$
with scaling dimension $-1/5$. In this latter case the general perturbed
action is \[
{\mathcal{A}}_{\lambda,\Phi(h)}={\mathcal{A}}_{\Phi}+\lambda\int\limits _{-\infty}^{\infty}dy\int\limits _{-\infty}^{0}dx\phi(x,y)+h\int\limits _{-\infty}^{\infty}dy\varphi(y),\]
 where ${\mathcal{A}}_{\Phi}$ denotes the action for ${\mathcal{M}}(2/5)$
with the $\Phi$ boundary condition imposed at $x=0$, and $\lambda$
and $h$ denote the bulk and boundary couplings respectively. The
action of ${\mathcal{A}}_{\lambda,\textrm{1}}$ is similar, but the
last term on the right hand side is missing. If $\lambda>0$ then
in all cases the bulk behaviour is described by an integrable massive
theory having only a single particle type with the following S matrix
\cite{CM}: \[
S(\theta)=-\left(\frac{1}{3}\right)\left(\frac{2}{3}\right)=-\left[\frac{1}{3}\right]\quad;\quad(x)=\frac{\sinh\left(\frac{\theta}{2}+\frac{i\pi x}{2}\right)}{\sinh\left(\frac{\theta}{2}-\frac{i\pi x}{2}\right)}.\]
 The pole at $\theta=\frac{2\pi i}{3}$ corresponds to the {}``$\varphi^{3}$
property'', i.e. the particle appears as a bound state of itself.
The minimal bulk two particle form factor which has only a zero at
$\theta=0$ and a pole at $\theta=\frac{2\pi i}{3}$ in the strip
$0\leq\Im m(\theta)<\pi$ has the form \cite{Z1}: \[
f(\theta)=\frac{y-2}{y+1}v(i\pi-\theta)v(-i\pi+\theta)\quad,\quad y=e^{\theta}+e^{-\theta},\]
 where \[
v(\theta)=\exp\left\{ 2\int_{0}^{\infty}\frac{dt}{t}e^{i\frac{\theta t}{\pi}}\frac{\sinh\frac{t}{2}\sinh\frac{t}{3}\sinh\frac{t}{6}}{\sinh^{2}t}\right\} .\]

In the boundary theory with the perturbed $\Phi$ boundary, the reflection
amplitude of the particle depends on the strength of the coupling
constant of the boundary field as \cite{DPTW1} \[
R(\theta)_{\Phi}=R_{0}(\theta)R(b,\theta)=\left(\frac{1}{2}\right)\left(\frac{1}{6}\right)\left(-\frac{2}{3}\right)\left[\frac{b+1}{6}\right]\left[\frac{b-1}{6}\right],\]
 where the dimensionless parameter $b$ is related to the dimensionful
$h$ as \[
h(b)=\sin\left(\bigl(b+\frac{1}{2}\bigr)\frac{\pi}{5}\right)m(\lambda)^{6/5}h_{crit},\qquad h_{crit}=-\pi^{\frac{3}{5}}2^{\frac{4}{5}}5^{\frac{1}{4}}\frac{\sin\frac{2\pi}{5}}{\sqrt{\Gamma(\frac{3}{5})\Gamma(\frac{4}{5})}}\left(\frac{\Gamma(\frac{3}{5})}{\Gamma(\frac{3}{5})}\right)^{\frac{6}{5}},\]
 and $m(\lambda)$ is the mass of the particle giving the overall
scale in the infrared description. In the case of the $\textrm{1}$
boundary the reflection amplitude is the parameter independent expression
\[
R(\theta)_{1}=\left(\frac{1}{2}\right)\left(\frac{1}{6}\right)\left(-\frac{2}{3}\right).\]
 Note that $R(\theta)_{1}$ is identical to $R(\theta)_{\Phi}$ at
$b=0$ and so both have a pole at $\theta=i\pi/2$ coming from the
$\left(\frac{1}{2}\right)$ block but their $g$ factors differ in
a sign \cite{DTWbct}.

\subsubsection{Lee-Yang model with perturbed $\Phi$ boundary }

We consider first the Lee-Yang model with perturbed $\Phi$ boundary.
The 1PFF corresponding to $R(\theta)_{\Phi}$ is chosen as \[
r(\theta)=\frac{i\sinh\theta}{(\sinh\theta-i\sin\frac{\pi(b+1)}{6})(\sinh\theta-i\sin\frac{\pi(b-1)}{6})}u(\theta),\]
 where \[
u(\theta)=\exp\left\{ \int_{0}^{\infty}\frac{dt}{t}\left[\frac{1}{\sinh\frac{t}{2}}-2\cosh\frac{t}{2}\cos\left[\left(\frac{i\pi}{2}-\theta\right)\frac{t}{\pi}\right]\frac{\sinh\frac{5t}{6}+\sinh\frac{t}{2}-\sinh\frac{t}{3}}{\sinh^{2}t}\right]\right\} .\]
 Note that $r\sim1$ at $y\rightarrow\infty$, and $r(\theta)$ satisfies
the $r(\theta+i\pi)=r(\theta)^{*}$ reality condition for real $\theta$.
The general $n$-particle form factors have the form \begin{equation}
F_{n}(\theta_{1},\dots,\theta_{n})=H_{n}Q_{n}(y_{1},\dots,y_{n})\prod_{i}\frac{r(\theta_{i})}{y_{i}}\prod_{i<j}\frac{f(\theta_{i}-\theta_{j})f(\theta_{i}+\theta_{j})}{y_{i}+y_{j}},\label{Ansatz}\end{equation}
 where we separated a normalizing factor $H_{n}$ from the polynomials
$Q_{n}$. The various $F_{n}$-s are related to each other by both
the kinematical and the dynamical singularity equations, since the
S-matrix is nontrivial and also has a $\varphi^{3}$ pole with $\Gamma=i2^{\frac{1}{2}}3^{\frac{1}{4}}$.
In addition, these $F_{n}$-s also have to satisfy the equation coming
from the residue of the pole at $\theta=i\pi/2$: \begin{equation}
-i\mathop{\textrm{Res}}_{\theta=i\pi/2}F_{n+1}(\theta,\theta_{1},\dots,\theta_{n})=\frac{g_{\Phi}}{2}\left(1-\prod\limits _{j=1}^{n}S(i\frac{\pi}{2}-\theta_{j})\right)F_{n}(\theta_{1},\dots,\theta_{n}).\label{eq:ipipol}\end{equation}
 Our strategy is to solve the recursion equations coming from the
first two conditions first and check whether the solutions also satisfy
the third requirement (\ref{eq:ipipol}). By choosing the normalizing
factors $H_{n}$ and introducing the useful quantities $\beta_{k}$
\begin{equation}
H_{n}=N\left(\frac{i3^{\frac{1}{4}}}{2^{\frac{1}{2}}v(0)}\right)^{n}\qquad\beta_{k}(b)=2\cos\frac{\pi}{6}(b+k),\quad k\in\mathrm{Z},\label{eq:Hn_betak}\end{equation}
 the overall normalization $N$ drops out and the recursion equations
coming from the dynamical (resp. kinematical) singularities read\begin{eqnarray}
Q_{2}(y_{+},y_{-}) & = & (y^{2}-\beta_{-3}^{2})Q_{1}(y),\nonumber \\
Q_{n+2}(y_{+},y_{-},y_{1},\dots,y_{n}) & = & Q_{n+1}(y,y_{1},\dots,y_{n})\,(y^{2}-\beta_{-3}^{2})\prod_{i=1}^{n}(y+y_{i}),\quad n>0;\label{eq:rec2a}\end{eqnarray}
\begin{equation}
Q_{n+2}(-y,y,y_{1},\dots,y_{n})=Q_{n}(y_{1},\dots,y_{n})\,(y^{2}-\beta_{-1}^{2})(y^{2}-\beta_{1}^{2})\, P_{n},\label{eq:rec2b}\end{equation}
 where \begin{equation}
P_{n}=\frac{1}{2(y_{+}-y_{-})}\left[\prod_{i=1}^{n}(y_{i}-y_{-})(y_{i}+y_{+})-\prod_{i=1}^{n}(y_{i}+y_{-})(y_{i}-y_{+})\right],\label{eq:Pequ}\end{equation}
 and \begin{equation}
y_{+}=\omega x+\omega^{-1}x^{-1};\quad y_{-}=\omega^{-1}x+\omega x^{-1};\quad x=e^{\theta};\quad\omega=e^{i\frac{\pi}{3}},\quad y=x+x^{-1}.\label{eq:ypmeq}\end{equation}
Next we present the minimal solution of these recursion equations
up to $n=3$. The solution is called minimal, if the leading overall
degrees of the $F_{n}$-s in all of the $y$ variables are the smallest
possible ones. Of course the solution also depends on the input function
$Q_{1}(y_{1})$. Since $F_{1}$ can have no pole at $\theta=i\pi/2$
while $r(\theta)/y$ has one, $Q_{1}$ must be chosen to cancel this
pole; the choice with the minimal degree is $Q_{1}(y_{1})=y_{1}=\sigma_{1}(y_{1})$.
Using this as input, we find from (\ref{eq:rec2a} ,\ref{eq:rec2b})
the \textsl{unique} solution \begin{eqnarray}
Q_{1}(y_{1})=\sigma_{1},\qquad\qquad\qquad\quad Q_{2}(y_{1},y_{2})=\sigma_{1}(\sigma_{2}+3-\beta_{-3}^{2}),\nonumber \\
Q_{3}(y_{1},y_{2},y_{3})=\sigma_{1}\left[\sigma_{1}(\sigma_{2}+\beta_{-1}^{2})(\sigma_{2}+\beta_{1}^{2})-(\sigma_{2}+3)(\sigma_{1}\sigma_{2}-\sigma_{3})\right].\label{eq:lysol1}\end{eqnarray}
 The remarkable property of this solution is that it contains no free
parameters. A simple counting of the various powers shows that the
leading overall degree of $F_{1}$, $F_{2}$, and $F_{3}$ vanish. 

To check eq.(\ref{eq:ipipol}) we need the following relations following
from the explicit solution (\ref{eq:lysol1}) and from the various
identities among the $\beta_{k}$-s: \begin{eqnarray*}
Q_{2}(0,y_{2}) & = & \sigma_{1}(y_{2})(3-\beta_{-3}^{2}),\\
Q_{3}(0,y_{2},y_{3}) & = & \beta_{1}\beta_{-1}\sigma_{1}(y_{2},y_{3})Q_{2}(y_{2},y_{3})=(3-\beta_{-3}^{2})\sigma_{1}(y_{2},y_{3})Q_{2}(y_{2},y_{3}).\end{eqnarray*}
 Indeed using them in eq.(\ref{eq:ipipol}) leads to a consistency
condition on the ratio of the $H_{n}$-s: \[
\frac{H_{n+1}}{H_{n}}r(i\frac{\pi}{2})(3-\beta_{-3}^{2})=-2i\sqrt{3}g_{\Phi},\qquad n=1,2.\]
 Since \[
r(i\frac{\pi}{2})=\frac{4u(i\frac{\pi}{2})}{i(\sqrt{3}-\beta_{-3})^{2}},\quad\textrm{and}\quad g_{\Phi}=i2(3)^{1/4}(2-\sqrt{3})^{1/2}\frac{\sqrt{3}+\beta_{-3}}{\sqrt{3}-\beta_{-3}},\]
 the $b$ dependence cancels from the consistency condition and using
the actual form of the $H_{n}$-s leads to \[
\frac{u(i\frac{\pi}{2})}{\sqrt{2}v(0)}=\sqrt{3}\,(2-\sqrt{3})^{1/2},\]
 which we checked numerically up to 7 digits.

To test these form factors numerically against the predictions of
conformal field theory, we take the spectral representation of the
boundary two-point function \begin{eqnarray*}
\langle0\vert\mathcal{O}(t)\mathcal{O}(0)\vert0\rangle & = & \sum_{n=0}^{\infty}\int_{0}^{\infty}\frac{d\theta_{1}\dots d\theta_{n}}{n!(2\pi)^{n}}F_{n}^{\mathcal{O}}(\theta_{1},\dots,\theta_{n})^{+}F_{n}^{\mathcal{O}}(\theta_{1},\dots,\theta_{n})e^{-mt\sum_{i=1}^{n}\cosh\theta_{i}}\\
 & = & \sum_{n=0}^{\infty}(-)^{n}\int_{0}^{\infty}\frac{d\theta_{1}\dots d\theta_{n}}{n!(2\pi)^{n}}\vert F_{n}^{\mathcal{O}}(\theta_{1},\dots,\theta_{n})\vert^{2}e^{-mt\sum_{i=1}^{n}\cosh\theta_{i}}\end{eqnarray*}
 which we truncate to the first few terms in the boundary form factor
expansion. Since the minimal solution of the form factor problem has
the mildest UV behaviour it is natural to assume, that in the UV it
corresponds to the boundary field $\varphi$. Therefore $\langle0\vert\mathcal{O}(t)\mathcal{O}(0)\vert0\rangle$
obtained from the FF expansion must be compared to the short distance
expansion: \[
\langle0\vert m^{\frac{1}{5}}\varphi(t)m^{\frac{1}{5}}\varphi(0)\vert0\rangle=-(mt)^{\frac{2}{5}}+(mt)^{\frac{1}{5}}C_{\varphi\varphi}^{\varphi}\langle m^{\frac{1}{5}}\varphi\rangle+\dots\]
where appropriate powers of $m$ were inserted to make the expression
dimensionless and the fusion coefficient is \[
C_{\varphi\varphi}^{\varphi}=-\sqrt{\frac{1+\sqrt{5}}{2}}\sqrt{\frac{\Gamma(\frac{1}{5})\Gamma(\frac{6}{5})}{\Gamma(\frac{3}{5})\Gamma(\frac{4}{5})}}\]
while the ($b$-dependent, dimensionless) vacuum expectation value
\[
\langle m^{\frac{1}{5}}\varphi\rangle=-\frac{5}{6h_{crit}}\frac{\cos(\frac{\pi b}{6})}{\cos(\frac{\pi}{10}(2b+1))}\]
 is given explicitly in \cite{DPTW1}. In analogy with the bulk case
\cite{DPTW1pt} we choose the normalization factor $N$ in (\ref{eq:Hn_betak})
as the vacuum expectation value of the boundary field \begin{equation}
N=\langle m^{\frac{1}{5}}\varphi\rangle\label{eq:LYnorm}\end{equation}
 With this choice the boundary form factor expansion gives \begin{eqnarray*}
\langle0\vert\mathcal{O}(t)\mathcal{O}(0)\vert0\rangle & = & \vert F_{0}^{\mathcal{O}}\vert^{2}-\int_{0}^{\infty}\frac{d\theta}{2\pi}\vert F_{1}^{\mathcal{O}}\vert^{2}e^{-mt\cosh\theta}\\
 &  & +\int_{0}^{\infty}\frac{d\theta_{1}d\theta_{2}}{2(2\pi)^{2}}\vert F_{2}^{\mathcal{O}}(\theta_{1},\theta_{2})\vert^{2}e^{-mt(\cosh\theta_{1}+\cosh\theta_{2})}\\
 &  & -\int_{0}^{\infty}\frac{d\theta_{1}d\theta_{2}d\theta_{3}}{6(2\pi)^{3}}\vert F_{3}^{\mathcal{O}}(\theta_{1},\theta_{2},\theta_{3})\vert^{2}e^{-mt(\cosh\theta_{1}+\cosh\theta_{2}+\cosh\theta_{3})}+\dots\end{eqnarray*}
The two expansions are compared on the next figure 

\begin{center}\includegraphics[%
  width=15cm]{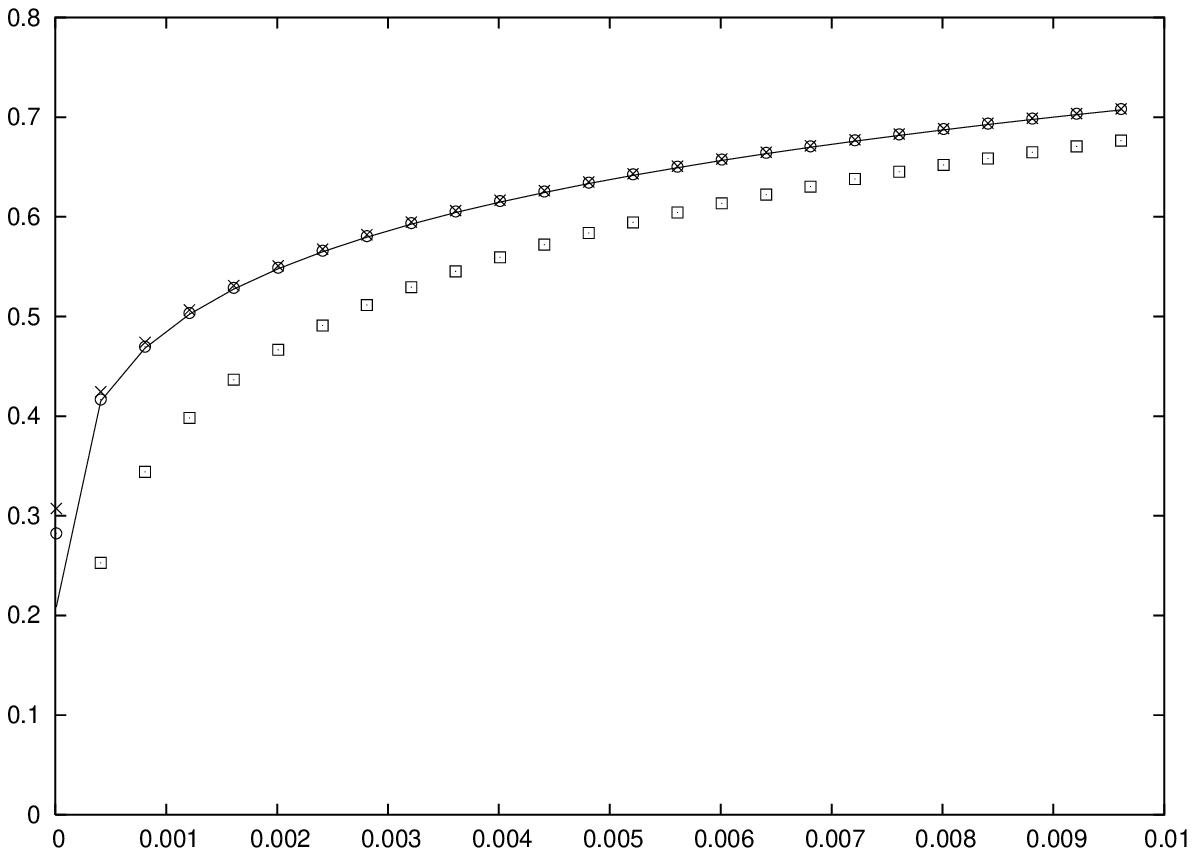}\end{center}

\noindent where $b=-1.05$ and the dimensionless correlation function
is plotted against $mt$. The predicted UV behaviour is given by the
continuous line and the numerically determined form factor expansion
truncated at 1, 2 and 3 particle intermediate states is denoted by
the symbols $\square$, $\times$ and $\circ$, respectively. 

We checked that the agreement between the form factor expansion truncated
at three particles and the UV CFT prediction holds for various values
of the parameter $b$: indeed as we change $b$ the two curves move
together. The agreement above also confirms our choice (\ref{eq:LYnorm})
for the normalization of the form factors. Based on all these we associate
the boundary operator corresponding to the minimal solution of the
form factor axioms to the one, that in the UV limit becomes the $\varphi$
field of the boundary Lee-Yang model.

\subsubsection{Lee-Yang model with the $\textrm{1}$ boundary}

The 1PFF corresponding to the parameter free $R(\theta)_{1}$ is chosen
as \[
r_{1}(\theta)=i\sinh\theta\,\, u(\theta),\]
 where $u(\theta)$ is the same as in the previous subsection. Note
that $r_{1}$ also satisfies $r_{1}(\theta+i\pi)=r_{1}(\theta)^{*}$
but its asymptotic behaviour $r_{1}\sim y^{2}$ at $y\rightarrow\infty$
is different from that of the $r$ in the previous case. Since $R(\theta)_{1}$
also has a pole at $\theta=i\pi/2$ we introduce a similar Ansatz
as in (\ref{Ansatz}) \begin{equation}
F_{n}(\theta_{1},\dots,\theta_{n})=H_{n}^{1}Q_{n}(y_{1},\dots,y_{n})\prod_{i}\frac{r_{1}(\theta_{i})}{y_{i}}\prod_{i<j}\frac{f(\theta_{i}-\theta_{j})f(\theta_{i}+\theta_{j})}{y_{i}+y_{j}},\label{Ansatz1}\end{equation}
with $H_{n}^{1}=4^{n}H_{n}$ where $H_{n}$ is the same as in (\ref{eq:Hn_betak}).
Then one finds the following recursion equations for the $Q_{n}$-s
from the dynamical (resp. kinematical) singularity equations: \begin{eqnarray}
Q_{2}(y_{+},y_{-}) & = & Q_{1}(y),\nonumber \\
Q_{n+2}(y_{+},y_{-},y_{1},\dots,y_{n}) & = & Q_{n+1}(y,y_{1},\dots,y_{n})\,\prod_{i=1}^{n}(y+y_{i}),\quad n>0;\label{eq:rec21a}\end{eqnarray}
\begin{equation}
Q_{n+2}(-y,y,y_{1},\dots,y_{n})=Q_{n}(y_{1},\dots,y_{n})\, P_{n},\label{eq:rec21b}\end{equation}
 where the various symbols are the same as in eq.(\ref{eq:rec2a},\ref{eq:rec2b}).
Up to $n=4$ the \textsl{unique} minimal solution of these recursion
equations with the input $Q_{1}(y_{1})=\sigma_{1}$ is \[
Q_{1}(y_{1})=\sigma_{1},\quad Q_{2}(y_{1},y_{2})\sim\sigma_{1},\quad Q_{3}(y_{1},y_{2},y_{3})\sim\sigma_{1}^{2},\quad Q_{4}(y_{1},y_{2},y_{3},y_{4})\sim\sigma_{1}^{2}(\sigma_{2}+3).\]
 It is easy to show that the leading overall degree of the first four
form factors $F_{1},\dots,F_{4}$ is two. This indicates that the
operator that corresponds to this set is different from the one encountered
in the case of the perturbed $\Phi$ boundary. Therefore in the conformal
limit this operator is different from the $\varphi$ field and this
is in accord with the fact that only the identity operator and its
descendents can live on the conformal boundary condition $1$. Based
on the asymptotics of the form factors for large $\theta$ the corresponding
operator has ultraviolet dimension $2$ and can be identified with
the unique such operator in the conformal vacuum module which is the
$L_{-2}$ descendent of the identity. \emph{}This identification is
further confirmed by comparing the numerically obtained truncated
form factor expansion to the conformal two-point function. 

Since \[
r_{1}(i\frac{\pi}{2})=-u(i\frac{\pi}{2}),\quad\textrm{and}\quad g_{1}=-i2(3)^{1/4}(2-\sqrt{3})^{1/2},\]
 (see also \cite{DTWbct}) one can readily show that these four form
factors also satisfy the equation coming from the residue of the pole
at $\theta=i\pi/2$.

\subsection{The boundary sinh-Gordon model}

The sinh-Gordon theory in the bulk is defined by the Lagrangian%
\footnote{Note that the parameter $b$ is used here with a different meaning
compared to the former case of the boundary Lee-Yang model. %
}:\[
\mathcal{L}=\frac{1}{2}(\partial_{\mu}\Phi)^{2}-\frac{m^{2}}{b^{2}}(\cosh b\Phi-1)\]
It can be considered as the analytic continuation of the sine-Gordon
model for imaginary coupling $\beta=ib$. The S-matrix of the model
is \[
S(\theta)=-\left(1+\frac{B}{2}\right)\left(-\frac{B}{2}\right)=-\left[-\frac{B}{2}\right]\qquad;\quad B=\frac{2b^{2}}{8\pi+b^{2}}\]
The minimal bulk two particle form factor belonging to this S-matrix
is \cite{FMS} \[
f(\theta)=\mathcal{N}\exp\left[8\int_{0}^{\infty}\frac{dx}{x}\sin^{2}\left(\frac{x(i\pi-\theta)}{2\pi}\right)\frac{\sinh\frac{xB}{4}\sinh(1-\frac{B}{2})\frac{x}{2}\sinh\frac{x}{2}}{\sinh^{2}x}\right],\]
 and it satisfies \begin{equation}
f(\theta)f(\theta+i\pi)=\frac{\sinh\theta}{\sinh\theta+i\sin\frac{\pi B}{2}}\,\,.\label{eq:ffip}\end{equation}
The sinh-Gordon theory can be restricted to the negative half-line,
but the integrability is maintained only by imposing either the Dirichlet\[
\Phi(0,t)=\Phi_{0}^{D}\]
or the two parameter family of perturbed Neumann\[
V_{B}(\Phi(0,t))=M_{0}\cosh\left(\frac{b}{2}(\Phi(0,t)-\Phi_{0})\right)-M_{0}\]
boundary conditions. The latter interpolates between the Neumann and
the Dirichlet boundary conditions, since for $M_{0}=0$ we recover
the Neumann, while for $M_{0}\to\infty$ the Dirichlet boundary condition
with $\Phi_{0}^{D}=\Phi_{0}$. The reflection factor which depends
on two continuous parameters can be written as \[
R(\theta)=R_{0}(\theta)R(E,F,\theta)=\left(\frac{1}{2}\right)\left(\frac{1}{2}+\frac{B}{4}\right)\left(-1-\frac{B}{4}\right)\left[\frac{E-1}{2}\right]\left[\frac{F-1}{2}\right]\]
 in terms of the parameterization used in \cite{shGEF}. They are
related to the parameters of the Lagrangian as \begin{eqnarray*}
\cos\frac{E}{16}(b^{2}+8\pi)\cos\frac{F}{16}(b^{2}+8\pi) & = & \frac{M_{0}}{M_{crit}}\cosh\frac{b\Phi_{0}}{2}\\
\sin\frac{E}{16}(b^{2}+8\pi)\sin\frac{F}{16}(b^{2}+8\pi) & =- & \frac{M_{0}}{M_{crit}}\sinh\frac{b\Phi_{0}}{2}\end{eqnarray*}
where $M_{crit}=m\sqrt{\frac{2}{b^{2}\sinh(b^{2}/8)}}$. Note that
for generic values of the parameters ($E\ne0$, $F\ne0$) this reflection
factor has a pole at $\theta=i\pi/2$ coming from the $\left(\frac{1}{2}\right)$
factor. Imposing Dirichlet boundary condition instead of the general
one corresponds to removing the $F$ dependent factor from $R(\theta)$.
Then the remaining parameter $E$ is related to the $\Phi_{0}$ boundary
value of the sinh-Gordon field as $E=i8b\Phi_{0}/(b^{2}+8\pi)$.

\subsubsection{Sinh-Gordon model with $\Phi_{0}=0$ Dirichlet b.c.}

This case is interesting because $E=0$ implies that the pole at $\theta=i\pi/2$
is \textsl{absent} in this case. Therefore the equation coming from
the residue of this pole is also absent and the form factors are less
restricted. The 1PFF corresponding to the reflection amplitude on
the $E=0$ Dirichlet boundary is \[
r_{0}(\theta)=\frac{\sinh\theta}{\sinh\theta+i}u(\theta,B),\]
 where \[
u(\theta,B)=\exp\left[-2\int_{0}^{\infty}\frac{dx}{x}\left[\cos(\frac{i\pi}{2}-\theta)\frac{x}{\pi}-1\right]\frac{\cosh\frac{x}{2}}{\sinh^{2}x}\left(\sinh\frac{xB}{4}+\sinh(1-\frac{B}{2})\frac{x}{2}+\sinh\frac{x}{2}\right)\right].\]
 Note that $r_{0}\sim y$ at $y\rightarrow\infty$ and has no pole
at $\theta=i\pi/2$. At $B=0$ - which corresponds to a free theory
- $u(\theta,0)$ can be integrated explicitly yielding $r_{0}(\theta)|_{B=0}=(-i\sinh\theta)/2$;
and this, apart from a trivial normalizational phase coincides with
the 1PFF for a free scalar with Dirichlet b.c. (As discussed in Section
3.1.1 in this case $\partial_{x}\Phi(0,0)$ is the operator having
one particle matrix element only). 

We write the $n$ particle form factors in the general form: \[
F_{n}(\theta_{1},\dots,\theta_{n})=H_{n}Q_{n}(y_{1},\dots,y_{n})\prod_{i}r_{0}(\theta_{i})\prod_{i<j}\frac{f(\theta_{i}-\theta_{j})f(\theta_{i}+\theta_{j})}{y_{i}+y_{j}}.\]
 Since there is no self fusing pole in the $S$ matrix of the sinh-Gordon
model, the $F_{n}$-s are related only by the kinematical singularity
equations. Choosing the ratio of the $H_{n}$-s appropriately the
recursion equations originating from here take the form: \[
Q_{n+2}(-y,y,y_{1},\dots,y_{n})=-Q_{n}(y_{1},\dots,y_{n})\, P_{n},\]
 where $P_{n}$ is given by eq.(\ref{eq:Pequ},\ref{eq:ypmeq}) with
$\omega=e^{i\pi\frac{B}{2}}$.

As $r_{0}$ has no pole at $\theta=i\pi/2$ , one can have a minimal
solution of this recursion equations starting with $Q_{1}=1$ which
has non vanishing form factors for \emph{odd} particle numbers. We
calculated up to $n=5$ and found that the solution is uniquely given
by \[
Q_{1}(y_{1})=1,\qquad Q_{3}(y_{1},y_{2},y_{3})=-\sigma_{1},\]
 \[
Q_{5}(y_{1},\dots,y_{5})=\sigma_{1}[\sigma_{3}\sigma_{2}-(\omega+\omega^{-1})^{2}\sigma_{5}+(\omega-\omega^{-1})^{4}\sigma_{1}-(\omega-\omega^{-1})^{2}(\sigma_{3}+\sigma_{1}\sigma_{2})],\]
 with all the $F_{1}$, $F_{3}$ and $F_{5}$ form factors having
leading overall degree one. There is a unique local operator with
this property, namely $\partial_{x}\Phi$. 

Of course one can also find non vanishing solutions with \emph{even}
particle numbers also starting with a non trivial $Q_{2}$. Since
$F_{2}$ can have no kinematical singularity, the minimal choice is
$Q_{2}(y_{1},y_{2})=\sigma_{1}$. With this input we obtained again
a unique solution \[
Q_{2}(y_{1},y_{2})=\sigma_{1},\qquad Q_{4}(y_{1},\dots,y_{4})=\sigma_{1}^{2}(\sigma_{2}-(\omega-\omega^{-1})^{2}),\]
 where both the $F_{2}$ and the $F_{4}$ have leading overall degree
two. 

For $\Phi_{0}=0$ ($E=0$) the $\Phi\rightarrow-\Phi$ reflection
symmetry of the bulk sinh-Gordon model survives also in the boundary
theory. Therefore the boundary operators can be classified as even
or odd ones, having only non-vanishing even or odd particle form factors,
respectively. Thus the second form factor family can be identified
with the operator $(\partial_{x}\Phi)^{2}$.

\subsubsection{Sinh-Gordon model with $\Phi_{0}\ne0$ Dirichlet b.c.}

For $\Phi_{0}\ne0$ ($E\ne0$) the reflection factor acquires a pole
at $\theta=i\pi/2$ due to the fact that the field has a nontrivial
vacuum configuration. At the same time the reflection symmetry of
the bulk sinh-Gordon model is violated in the boundary theory. Therefore
the boundary operators cannot be classified into representations of
this symmetry, and the equation coming from the residue of the pole
at $\theta=i\pi/2$ connects the form factors with even and odd particle
numbers. Note that now this equation plays an essential role as it
is the only one that relates these two sets of form factors to each
other.

The 1PFF corresponding to the reflection amplitude on the $E\ne0$
Dirichlet boundary is \[
r_{E}(\theta)=\frac{\sinh\theta}{\sinh\theta-i\sin\gamma}u(\theta,B)\quad,\qquad\qquad\gamma=\frac{\pi}{2}(E-1),\]
 where $u(\theta,B)$ is the same as in the previous case. \emph{}Note
that $r_{E}\sim y$ at $y\rightarrow\infty$. \emph{}

Writing the $n$ particle form factors in the general form \[
F_{n}(\theta_{1},\dots,\theta_{n})=H_{n}Q_{n}(y_{1},\dots,y_{n})\prod_{i}\frac{r_{E}(\theta_{i})}{y_{i}}\prod_{i<j}\frac{f(\theta_{i}-\theta_{j})f(\theta_{i}+\theta_{j})}{y_{i}+y_{j}},\]
and choosing the ratio of the $H_{n}$-s appropriately the recursion
equations originating from the kinematical singularity equation take
the form: \begin{equation}
Q_{n+2}(-y,y,y_{1},\dots,y_{n})=(y^{2}-4\cos^{2}\gamma)Q_{n}(y_{1},\dots,y_{n})\, P_{n},\label{eq:shGkin}\end{equation}
 where $P_{n}$ is given by eq.(\ref{eq:Pequ},\ref{eq:ypmeq}) with
$\omega=e^{i\pi\frac{B}{2}}$. Next we show how the equation coming
from the residue of the pole at $\theta=i\pi/2$: \begin{equation}
-i\mathop{\textrm{Res}}_{\theta=i\pi/2}F_{n+1}(\theta,\theta_{1},\dots,\theta_{n})=\frac{g_{E}}{2}\left(1-\prod\limits _{j=1}^{n}S(i\frac{\pi}{2}-\theta_{j})\right)F_{n}(\theta_{1},\dots,\theta_{n}),\label{eq:ipipol2}\end{equation}
 helps to eliminate the arbitrariness in the minimal solution of the
recursion equations.

$Q_{1}(y_{1})=\sigma_{1}$ is the minimal choice that guarantees that
$F_{1}$ has no pole at $\theta=i\pi/2$. Using this in the recursion
equation (\ref{eq:shGkin}) gives that the most general $Q_{3}$ has
the form: \[
Q_{3}(y_{1},y_{2},y_{3})=-\sigma_{1}^{2}(\sigma_{2}+4\cos^{2}\gamma)+(A+B\sigma_{1})(\sigma_{1}\sigma_{2}-\sigma_{3}),\]
 where $A$ and $B$ are arbitrary constants. Eq.(\ref{eq:ipipol2})
leads to the following relation between $Q_{3}$ and $Q_{2}$: \[
H_{3}r_{E}(i\frac{\pi}{2})Q_{3}(0,y_{2},y_{3})=g_{E}2\sin\frac{B\pi}{2}H_{2}\sigma_{1}(y_{2},y_{3})Q_{2}(y_{2},y_{3}).\]
 Since \[
Q_{3}(0,y_{2},y_{3})=\sigma_{1}\left\{ -\sigma_{1}(\sigma_{2}+4\cos^{2}\gamma)+(A+B\sigma_{1})\sigma_{2}\right\} ,\]
 the expression in the curly bracket should be proportional to $Q_{2}$.
This observation fixes the values of $A$ and $B$: $Q_{2}$ has to
be proportional to $\sigma_{1}$ to guarantee that $F_{2}$ has no
kinematical singularity and this requirement is met only if $A=0$,
while $F_{2}$ has a leading degree not exceeding that of $F_{1}$
and $F_{3}$ if $B=1$. Thus with these two requirements one obtains
a parameter free solution starting with $Q_{1}=\sigma_{1}$; up to
$n=4$ it has the form: \[
Q_{2}(y_{1},y_{2})\sim-4\cos^{2}\gamma\sigma_{1},\quad Q_{3}(y_{1},y_{2},y_{3})=-\sigma_{1}\sigma_{3}-4\cos^{2}\gamma\sigma_{1}^{2},\]
 \[
Q_{4}(y_{1},\dots,y_{4})\sim-4\cos^{2}\gamma(\sigma_{1}\sigma_{3}+4\cos^{2}\gamma\sigma_{1}^{2})(\sigma_{2}+4\sin^{2}\frac{\pi B}{2}).\]
 Note that both $Q_{2}$ and $Q_{4}$ vanish for $\gamma=-\pi/2$
($E=0$). Furthermore for $E=0$ one also has $Q_{1}/y_{1}=1$, and
$Q_{3}/(y_{1}y_{2}y_{3})=-\sigma_{1}$, thus the solution goes over
smoothly into the one with $E=0$. Since \[
r_{E}(i\frac{\pi}{2})=\frac{1}{1-\sin\gamma},\qquad{\textrm{and}}\qquad g_{E}=\frac{2(1+\cos\frac{\pi B}{4}+\sin\frac{\pi B}{4})}{\sqrt{\sin\frac{\pi B}{2}}}\frac{\cos\gamma}{1-\sin\gamma}\,,\]
 the $\gamma$ dependence cancels from the ratios of $H_{3}/H_{1}$
and $H_{4}/H_{2}$ when we use eq.(\ref{eq:ipipol2}) for $n=1,2,3$:
\[
-i\frac{H_{4}}{H_{2}}=-i\frac{H_{3}}{H_{1}}=\left(1+\cos\frac{\pi B}{4}+\sin\frac{\pi B}{4}\right)^{2}4\sin\frac{\pi B}{2}.\]
In the $b\to0$ limit these ratios vanish, therefore the higher form
factors decouple in accord with the fact that the kinematical singularity
axiom becomes trivial for the free field theory.

\section{Conclusion}

In this paper we treated the form factor bootstrap for boundary operators
in integrable boundary quantum field theory. Although there have been
earlier treatment of form factors for specific (mainly lattice) models
\cite{XXZ,XXZGen,SGff}, none of these has actually given a complete
formulation similar to the axiomatic approach by Smirnov for the bulk
case \cite{Smirnov}. The present work initiates an extension of this
axiomatic program to boundary fields. 

We have given a complete axiomatization of the properties of boundary
form factors, derived from first principles of quantum field theory
(unitarity and the boundary extension of the LSZ reduction formulae).
In particular, the axiom describing boundary kinematic singularities
is an entirely new result of this paper, as this has never been treated
before in any previous study. We have shown that these axioms are
consistent with many known aspects of integrable boundary field theory.
In particular, the relation between the residue of the reflection
factor at $i\pi/2$ and the one-particle contribution to the boundary
state, noted previously in the context of finite size effects, was
confirmed once more as a necessary condition for the consistency between
the boundary and the bulk kinematical axiom (the only exception to
this relation is when the bulk is free, but then the two axioms are
trivial). Therefore it seems that this particular relation is a consequence
of integrability and the existence of a nontrivial bulk scattering
matrix.

We then proceeded to give a systematic method to solve the boundary
form factor axioms for the case of diagonal scattering. The solution
is a natural generalization of the bulk case, but necessitates the
introduction of a minimal boundary form factor function in addition
to the already known minimal bulk form factor. The periodicity, permutation
and reflection axioms can then be solved by a general Ansatz, and
the residue axioms can be recast as recursion relations for certain
\emph{polynomial} functions which characterize the form factor solution
completely.

In particular, we treated the case of the free boson and the free
fermion (noncritical Ising model with boundary magnetic field), where
the polynomial solutions of the form factor axioms were shown to be
identical to the explicit solutions obtained from the field theory,
and it was also shown that the polynomial solutions of the form factor
bootstrap match the full boundary operator content expected from the
Lagrangian approach. 

As example for the interacting case, we first treated the Lee-Yang
model, where the boundary kinematical singularity axiom makes its
first appearance, and it is very important in order to distinguish
between boundary conditions that have different conformal limits.
We have also computed the spectral expansion of the two-point correlation
function for the operator with the lowest conformal dimension and
have shown that it matches perfectly with the ultraviolet expansion
of the same correlation function obtained from boundary conformal
field theory.

Our second interacting example is the sinh-Gordon model, with Dirichlet
boundary condition (an extension to the general case is in principle
straightforward, but we decided to treat only the Dirichlet case to
keep it short and simple). The boundary conditions with zero and with
nonzero value of the field on the boundary are differentiated again
by the boundary kinematical axiom, and we have shown that the results
of the boundary form factor bootstrap fit perfectly well with expectations
from the Lagrangian approach.

An open question is to find and classify non-minimal solutions of
the form factor equations and interpret them in terms of the local
boundary operator algebra of the underlying field theory, extending
the method presented for the bulk sinh-Gordon model in \cite{koubek_mussardo}.
In particular it is interesting to find out whether the counting of
operators in the conformal limit can be matched with the full set
of solutions in the interacting case. 

It is obvious that the results presented in this paper can be applied
directly to any integrable boundary quantum field theory for which
the factorized scattering theory is known, and that they formulate
a well-defined program to determine form factors and correlation functions
of boundary operators, similar to the approach used in the bulk case.
We have also shown how to solve the axioms for theories with diagonal
bulk and boundary scattering.

It is an interesting problem to extend these results to the case of
nondiagonal scattering (with boundary sine-Gordon theory as the most
prominent example). The extension of the axioms is straightforward:
they must be decorated by multiplet indices, just like in the bulk
case, although here we avoided to give this extension explicitly to
keep the exposition simple. However, solving them will probably encounter
much more difficulties, and just as in the bulk, new methods must
be devised for the task, like the boundary extension of the Lukyanov
free field representation in \cite{SGff}.

The comparison to the Lagrangian and perturbed conformal field theory
description would be greatly facilitated by establishing sum rules
for the spectral representation of the boundary correlators, similar
to the $c$-theorem \cite{c-th} and $\Delta$-theorem \cite{delta-th}
in the bulk case, and is one of the most important problems left open
by the present work. Another promising open direction is to consider
possible applications of boundary form factors and correlation functions
in the area of boundary quantum field theory and condensed matter.

\subsection*{Acknowledgments}

We are grateful to F.A. Smirnov and G. Watts for useful discussions.
This research was partially supported by the EC network {}``EUCLID'',
contract number HPRN-CT-2002-00325, and Hungarian research funds OTKA
T043582, K60040 and TS044839. GT was also supported by a Bolyai J\'anos
research scholarship.

\appendix

\section{Heuristic derivation of the FF axioms}

We present some heuristic arguments, along the lines of \cite{Smirnov},
for the derivation of boundary form factor axioms using the boundary
reduction formula \cite{BBT}. 

We analyze the analyticity properties of the form factor\[
F_{n}^{\mathcal{O}}:=F_{n}^{\mathcal{O}}(\theta_{1},\theta_{2},\dots,\theta_{n})=\langle0\vert\mathcal{O}(0)\vert\theta_{1},\theta_{2},\dots,\theta_{n}\rangle_{in}\]
 as a function of the variable $\theta_{1}$. 

We follow the notations of \cite{BBT}: The asymptotic creation/annihilation
operators can be expressed in terms of the free asymptotic fields
as \begin{eqnarray}
a_{in}(\theta) & = & 2i\int_{-\infty}^{0}dx\cos(k(\theta)x)e^{i\omega(\theta)t}{\mathop{\partial}^{\leftrightarrow}}_{t}\Phi_{in}(x,t)\label{aa+}\\
a_{in}^{+}(\theta) & = & -2i\int_{-\infty}^{0}dx\cos(k(\theta)x)e^{-i\omega(\theta)t}{\mathop{\partial}^{\leftrightarrow}}_{t}\Phi_{in}(x,t)\;.\nonumber \end{eqnarray}
 where $k(\theta)=m\sinh\theta$ and $\omega(\theta)=m\cosh\theta$.
The \emph{in} state is a free state and we have \begin{equation}
\langle0\vert\mathcal{O}(0)\vert\theta_{1},\theta_{2},\dots,\theta_{n}\rangle_{in}=\langle0\vert\mathcal{O}(0)a_{in}^{+}(\theta_{1})\vert\theta_{2},\dots,\theta_{n}\rangle_{in}\label{matel}\end{equation}
We use (\ref{aa+}) together with \[
\mathcal{O}(0)\Phi_{in}(x,t)=[\mathcal{O}(0),\Phi_{in}(x,t)]+\Phi_{in}(x,t)\mathcal{O}(0)\]
to obtain \begin{equation}
F_{n}^{\mathcal{O}}=\textrm{disc.}-2i\int_{-\infty}^{0}dx\cos(k(\theta_{1})x)e^{-i\omega(\theta_{1})t}{\mathop{\partial}^{\leftrightarrow}}_{t}\,\langle0\vert[\mathcal{O}(0),\Phi_{in}(x,t)]\vert\theta_{2},\dots,\theta_{n}\rangle_{in}\;.\label{disc1}\end{equation}
where the disconnected part is \[
\textrm{disc.}=\langle0\vert a_{in}^{+}(\theta_{1})\mathcal{O}(0)\vert\theta_{2},\dots,\theta_{n}\rangle_{in}=\langle0\vert a_{in}^{+}(\theta_{1})\vert0\rangle F_{n-1}^{\mathcal{O}}(\theta_{2},\dots,\theta_{n})\]
Note that in theories with nonzero vacuum expectation values of the
field $\Phi$ the matrix element $\langle0\vert a_{in}^{+}(\theta_{1})\vert0\rangle$
is nonzero and can be written as \[
\langle0\vert a_{in}^{+}(\theta_{1})\vert0\rangle=\frac{g}{2}2\pi\delta(\theta_{1}-\frac{i\pi}{2})\]
which corresponds to the one particle term in the boundary state in
the crossed channel \cite{GZ}. It was conjectured in \cite{DPTW1pt}
and later confirmed using TBA arguments \cite{BLusch} that the one
particle contribution to the boundary state has a coefficient equal
to $\frac{g}{2}$ rather than $g$ as suggested in \cite{GZ}. In
the channel we use here this translates directly into the equation
above. 

Supposing that the \emph{in} field can be expressed in terms of the
interacting field as $\Phi(x,t)\to Z^{1/2}\Phi_{in}(x,t)$ for $t\to-\infty$
and that $[\mathcal{O}(0),\Phi(x,0)]=0$, the connected part can be
written in the form \[
\textrm{conn.}=iZ^{-1/2}2\int_{-\infty}^{0}dx\int_{-\infty}^{0}dt\partial_{t}\biggl\{\cos(k(\theta_{1})x)e^{-i\omega(\theta_{1})t}{\mathop{\partial}^{\leftrightarrow}}_{t}\,\langle0\vert[\mathcal{O}(0),\Phi(x,t)]\vert\theta_{2},\dots,\theta_{n}\rangle_{in}\biggr\}\]
Performing the usual partial integration while taking care of the
surface term we obtain\begin{equation}
\textrm{conn.}=iZ^{-1/2}2\int d^{2}xe^{-i\omega(\theta_{1})t}\cos(k(\theta_{1})x)\Theta(-t)\langle0\vert[\mathcal{O}(0),J(x,t)]\vert\theta_{2},\dots,\theta_{n}\rangle_{in}\label{redrin}\end{equation}
where $J(x,t)=\{\square+m^{2}+\delta(x)\partial_{x}\}\Phi(x,t)$ and
the integration goes over the entire spacetime. The range of the integration
is the interior of the past light cone due to the presence of $\Theta(-t)$
and of the vanishing of $[\mathcal{O}(0),J(x,t)]$ on space-like intervals.
The analytic properties of the integral are determined by the exponent
for large negative times. The exponent decreases if $\Im m(\omega(\theta_{1}))>0$
thus the \emph{in} form factor ($\theta_{1}>\theta_{2}>\dots>\theta_{n}>0$)
allows an analytical continuation into the domain:\[
0<\Im m(\theta_{1})<\pi\]

Repeating the same procedure for the \emph{out} matrix elements \[
F_{n}^{\mathcal{O}}(-\theta_{1},-\theta_{2},\dots,-\theta_{n})=\langle0\vert\mathcal{O}(0)\vert-\theta_{1},-\theta_{2},\dots,-\theta_{n}\rangle_{out}\]
we obtain the domain of analytical continuation: $0<\Im m(-\theta_{1})<\pi$. 

To derive the crossing relation we consider the following matrix element\[
F_{1n-1}^{\mathcal{O}}:=F_{1n-1}^{\mathcal{O}}(\theta_{1}\vert\theta_{2},\dots,\theta_{n})=\,_{in}\langle\theta_{1}\vert\mathcal{O}(0)\vert\theta_{2},\dots,\theta_{n}\rangle_{in}\]
Applying the reduction formula to the particle with rapidity $\theta_{1}$
(\ref{aa+}) we obtain\[
F_{1n-1}^{\mathcal{O}}=\textrm{disc}-2i\int_{-\infty}^{0}dx\cos(k(\theta_{1})x)e^{i\omega(\theta_{1})t}{\mathop{\partial}^{\leftrightarrow}}_{t}\,\langle0\vert[\mathcal{O}(0),\Phi_{in}(x,t)]\vert\theta_{2},\dots,\theta_{n}\rangle_{in}\]
where the disconnected part (supposing $\theta_{1}\geq\theta_{2}$)
is \[
\textrm{disc.}=\langle0\vert\mathcal{O}(0)a_{in}^{+}(\theta_{1})\vert\theta_{2},\dots,\theta_{n}\rangle_{in}=\,_{in}\langle\theta_{1}\vert\theta_{2}\rangle_{in}F_{n-2}^{\mathcal{O}}(\theta_{3},\dots,\theta_{n})\]
Performing a partial integration as before the result for the connected
component is \begin{equation}
\textrm{conn.}=iZ^{-1/2}2\int d^{2}xe^{i\omega(\theta_{1})t}\cos(k(\theta_{1})x)\Theta(-t)\langle0\vert[\mathcal{O}(0),J(x,t)]\vert\theta_{2},\dots,\theta_{n}\rangle_{in}\label{redlin}\end{equation}
 which has an analytic continuation for \[
-\pi<\Im m(\theta_{1})<0\]
 Comparing (\ref{redrin}) with (\ref{redlin}) and using that $m\cosh(\theta_{1}+i\pi)=-m\cosh\theta_{1}$
the crossing relation (\ref{eq:crossing}) is proved. Similar result
can be obtained for an \emph{out} state and the $-\theta_{1}<-\theta_{2}<\dots<-\theta_{n}<0$
range of the parameters. 

The reflection property (Axiom II) can be shown by considering \[
F_{n}^{\mathcal{O}}(\theta_{1},\theta_{2},\dots,\theta_{n})=\langle0\vert\mathcal{O}(0)\vert\theta_{1},\theta_{2},\dots,\theta_{n}\rangle_{in}\]
 and crossing all particles except the one with rapidity $\theta_{n}$
to the left. Now inserting a complete set of \emph{out} states we
have \[
_{in}\langle\dots\vert\mathcal{O}(0)\vert\theta_{n}\rangle_{in}=\sum_{out}\,_{in}\langle\dots\vert A(0)\vert n\rangle_{out}\,_{out}\langle n\vert\theta_{n}\rangle_{in}\]
where only the first two terms are nonzero: \[
_{in}\langle\dots\vert\mathcal{O}(0)\vert\theta_{n}\rangle_{in}=\,_{in}\langle\dots\vert A(0)\vert0\rangle\langle0\vert\theta_{n}\rangle_{in}+\sum_{\theta}\langle\dots\vert A(0)\vert\theta\rangle_{out}\,_{out}\langle\theta\vert\theta_{n}\rangle_{in}\]
The connected part gives the required $R$ factor while the disconnected
one combined with the disconnected part in (\ref{disc1}) and the
permutation property gives the boundary kinematical singularity. 

The permutation property in the bulk case is usually derived from
very similar argumentation we used above for showing the reflection
property. Note, however, that the same result can be obtained from
the analysis of the singularity structure of the Green functions:
the part, which is responsible for the discontinuity in the form factor
by changing two neighboring rapidities, is related to the bulk $S$-matrix.
The permutation property in the boundary case (Axiom I) can be derived
only from the second approach, namely from a detailed investigation
of the singularity structure of the Green functions. By extending
the result on the two point function in \cite{BBT} one can show that
multi point functions have momentum preserving parts identical to
their bulk counterparts and exactly these parts contribute only, when
two neighboring (both positive) rapidities are changed, and cause
the same discontinuity in the form factor we met in the bulk case.

The kinematical singularity equation (Axiom IV) can be obtained (using
the permutation and reflection axioms) from the analysis of the disconnected
components in the crossing relations as obtained for the \emph{in}
and for the \emph{out} states: \[
F_{1n-1}^{\mathcal{O}}(\pm\theta_{1}\vert\pm\theta_{2},\dots,\pm\theta_{n})=F_{n}^{\mathcal{O}}(i\pi\pm\theta_{1},\pm\theta_{2},\dots,\pm\theta_{n})+2\pi\delta(\theta_{1}-\theta_{2})F_{n-2}^{\mathcal{O}}(\pm\theta_{3},\dots,\pm\theta_{n})\]

Although our derivation of the boundary form factor axioms is heuristic
to some extent we expect that the formulation of the same ideas in
a rigorously defined quantum field theoretical framework would lead
to the proper and mathematically founded derivation (but note that
this has not been performed yet in the bulk case either).

\section{Formal derivation of the FF axioms }

Here we show how the boundary form factor axioms can be formally derived
from a boundary analogue of the Faddeev-Zamolodchikov algebra. 

In the bulk case one introduces creation $Z^{*}(\theta)$ and annihilation
$Z(\theta)$ operators corresponding to asymptotic states. They are
defined for real rapidities $\theta\in R$ and satisfy the following
defining relations \begin{eqnarray}
Z^{*}(\theta_{1})Z^{*}(\theta_{2}) & = & S(\theta_{1}-\theta_{2})Z^{*}(\theta_{2})Z^{*}(\theta_{1})\nonumber \\
Z(\theta_{1})Z(\theta_{2}) & = & S(\theta_{1}-\theta_{2})Z(\theta_{2})Z(\theta_{1})\nonumber \\
Z(\theta_{1})Z^{*}(\theta_{2}) & = & S(\theta_{2}-\theta_{1})Z^{*}(\theta_{2})Z(\theta_{1})+2\pi\delta(\theta_{1}-\theta_{2})\label{eq:ZFDEF}\end{eqnarray}
 Extending $Z,Z^{*}$ to imaginary rapidities (treating $\theta$
as a $2\pi i$ periodic complex variable) we encounter singularities
in their products at $\theta_{1}=\theta_{2}\pm i\pi$ with residues\[
-i\mathop{\textrm{Res}}_{\theta_{1}=\theta_{2}+i\pi}Z^{*}(\theta_{1})Z^{*}(\theta_{2})=1\]
and \[
-i\mathop{\textrm{Res}}_{\theta_{1}=\theta_{2}+i\pi}Z(\theta_{1})Z(\theta_{2})=1\]
These can be formulated by postulating the crossing property \[
Z(\theta)=Z^{*}(\theta+i\pi)\]
 and taking into account the defining relations (\ref{eq:ZFDEF}).
We note that using this identification all of the defining relations
(\ref{eq:ZFDEF}) can be combined into a single one\[
Z^{*}(\theta_{1})Z^{*}(\theta_{2})=S(\theta_{1}-\theta_{2})Z^{*}(\theta_{2})Z^{*}(\theta_{1})+2\pi\delta(\theta_{1}-\theta_{2}-i\pi)\]
 In the boundary case the generators $Z,Z^{*}$ are defined only for
positive values of the rapidity arguments and additionally two new
formal generators are introduced creating the boundary vacuum as follows\[
\vert0\rangle_{B}=B^{*}\vert0\rangle\quad,\qquad\,_{B}\langle0\vert=\langle0\vert B\]
We introduce two new relations \[
Z^{*}(\theta)B^{*}=R(\theta)Z^{*}(-\theta)B^{*}\]
and \[
BZ(\theta)=BZ(-\theta)R(-\theta)\]
which describe how we can extend the generators for negative rapidities.
By analytically continuing in the rapidity again we have singularities
in the operator products \[
-i\mathop{\textrm{Res}}_{\theta=i\frac{\pi}{2}}Z^{*}(\theta)B^{*}=\frac{g}{2}B^{*}\]
 and \[
i\mathop{\textrm{Res}}_{\theta=-i\frac{\pi}{2}}BZ(\theta)=\frac{g}{2}B\]
They would correspond to particles with real rapidity in the crossed
channel (exchanging time and space coordinates). These new relations
can again be summarized in a single one\[
Z^{*}(\theta)B^{*}=R(\theta)Z^{*}(-\theta)B^{*}+2\pi\delta(\theta-\frac{i\pi}{2})\frac{g}{2}B^{*}\]
together with its formal conjugate \[
BZ(\theta)=BZ(-\theta)R(-\theta)+2\pi\delta(\theta+\frac{i\pi}{2})\frac{g}{2}B\]
 We claim that representing the form factor of the boundary operator
$\mathcal{O}(0)$ as \[
F_{n}^{\mathcal{O}}(\theta_{1},\dots,\theta_{n})=\langle0\vert B\,\mathcal{O}(0)\, Z^{*}(\theta_{1})\dots Z^{*}(\theta_{n})B^{*}\vert0\rangle\]
and supposing locality \[
[\mathcal{O}(0),Z^{*}(\theta)]=0\]
we can recover all the non-singularity type form factor axioms immediately.
For deriving the singularity axioms we have to observe that singularity
appears not only from a single term. E.g. in the boundary kinematical
singularity axiom, the form factor $F_{n}^{\mathcal{O}}(\theta_{1},\dots,\theta_{n})$
exhibits a singularity in $\theta_{1}$ at $i\frac{\pi}{2}$ coming
from two places: the operator product of both $B$ and $B^{*}$ with
$Z^{*}(\theta_{1})$ is singular. Supposing that they appear in additive
terms of the form factor we can obtain the desired formula.

Finally, we note that formulating the boundary FZ algebra in the spirit
of \cite{Max} might lead to a more rigorous derivation of our axioms.


\begin{thebibliography}{10}
\bibitem{Sbstr}G. Mussardo\emph{, Phys.Rept}. \textbf{218}, (1992) 215-379. 
\bibitem{Sanal}P. E. Dorey, \emph{Exact S matrices}, Lecture Notes in Physics, Springer,
eds. Zal\'an Horv\'ath and L\'aszl\'o Palla, E\"otv\"os Summer
School in Physics: Conformal Field Theories and Integrable Models,
Budapest, Hungary, 13-18 Aug 1996. 
\bibitem{Smirnov}F.A. Smirnov: \texttt{Form-factors in completely integrable models
of quantum field theory}, \emph{Adv. Ser. Math. Phys}. \textbf{14}
(1992) 1-208. 
\bibitem{Arndt}H. Babujian, A. Fring, M. Karowski, A. Zapletal, \emph{Nucl.Phys.}
\textbf{B538} (1999) 535-586.
\bibitem{GZ}S. Ghoshal and A.B. Zamolodchikov, \emph{Int. J. Mod. Phys.} \textbf{A9}
(1994) 3841-3886 (Erratum-ibid. \textbf{A9} 4353), hep-th/9306002. 
\bibitem{Muss1pt}G. Mussardo, \emph{Spectral Representation of Correlation Functions
in two-dimensional Quantum Field Theories,} hep-th/9405128.
\bibitem{DPTW1pt}P. Dorey, M. Pillin, R. Tateo and G. Watts, \emph{Nucl. Phys.} \textbf{B594}
(2001) 625-659. 
\bibitem{cardy_mussardo}J.L. Cardy and G. Mussardo, \emph{Nucl. Phys.} \textbf{B340} (1990)
387-402.
\bibitem{BBT}Z. Bajnok, G. B\"{o}hm and G. Tak\'{a}cs, \emph{J. Phys.} \textbf{A35}
(2002) 9333-9342, hep-th/0207079.\\
 Z. Bajnok, G. B\"{o}hm and G. Tak\'{a}cs, \emph{Nucl. Phys.} \textbf{B682}
(2004) 585-617, hep-th/0309119. 
\bibitem{XXZ}M. Jimbo, R. Kedem, H. Konno, T. Miwa, R. Weston, \emph{Nucl. Phys.}
\textbf{B448} (1995) 429-456. 
\bibitem{XXZGen}Y.-H. Quano, \emph{Int.J.Mod.Phys.} \textbf{A15} (2000) 3699-3716,
\emph{J. Phys.} \textbf{A33} (2000) 8275, \emph{J.Phys.} \textbf{A34}
(2001) 8445-8464. 
\bibitem{SGff}B. Hou, K. Shi, Y. Wang, W.-l. Yang, \emph{Int. J. Mod. Phys.} \textbf{A12}
(1997) 1711-1741.
\bibitem{DTWbct}P. Dorey, R. Tateo and G. Watts, \emph{Phys. Lett.} \textbf{B448}
(1999) 249-256.
\bibitem{FMS}A. Fring, G. Mussardo, P. Simonetti, \emph{Nucl.Phys.} \textbf{B393}
(1993) 413-441. 
\bibitem{KW}M. Karowski and P. Weisz, \emph{Nucl. Phys.} \textbf{B139} (1978)
455-476. 
\bibitem{counting_ops}G. Delfino and G. Mussardo, \emph{Nucl. Phys.} \textbf{B455} (1995)
724-758, hep-th/9507010.
\bibitem{Z1}Al.B. Zamolodchikov, \emph{Nucl. Phys.} \textbf{B348} (1991) 619-641. 
\bibitem{DPTW1}P. Dorey, A. Pocklington, R. Tateo and G. Watts, \emph{Nucl. Phys.}
\textbf{B525} (1998) 641-663. 
\bibitem{CM}J.L. Cardy and G. Mussardo, \emph{Phys. Lett.} \textbf{B225} (1989)
275-278. 
\bibitem{shGEF}E. Corrigan, A. Taormina, \emph{J.Phys.} \textbf{A33} (2000) 8739.
\bibitem{koubek_mussardo}A. Koubek and G. Mussardo, \emph{Phys. Lett.} \textbf{B311} (1993)
193-201, hep-th/9306044.
\bibitem{c-th}A. B. Zamolodchikov, \emph{Pis'ma Zh Eksp. Theor. Fiz.} \textbf{43}
(1986) 565. (\emph{JETP Lett.} \textbf{43} (1986) 730.)
\bibitem{delta-th}G. Delfino, P. Simonetti and J-L. Cardy, \emph{Phys. Lett.} \textbf{B387}
(1996) 327-333.
\bibitem{BLusch}Z. Bajnok, L. Palla and G. Takács, \emph{Nucl. Phys.} \textbf{B716}
(2005) 519-542.
\bibitem{Max}M.R. Niedermaier, \emph{Nucl.Phys.} \textbf{B440} (1995) 603-646;
Erratum-ibid. \textbf{B456} (1995) 755.
\end{thebibliography}
\end{document}